\PassOptionsToPackage{table}{xcolor}

\pdfoutput=1
\documentclass[10pt,journal,compsoc]{IEEEtran}
\IEEEoverridecommandlockouts

\ifCLASSOPTIONcompsoc
  \usepackage[nocompress]{cite}
\else
  \usepackage{cite}
\fi

\ifCLASSINFOpdf
  \usepackage[pdftex]{graphicx}
\else
\fi
\usepackage[utf8]{inputenc}

\usepackage[utf8]{inputenc}
\usepackage{url}
\usepackage{changes}
\usepackage{graphicx}
\usepackage{subfigure}
\usepackage{listings}
\usepackage{caption}
\usepackage{xtab,booktabs}
\usepackage[bookmarks=false,hidelinks]{hyperref}
\usepackage{xspace}
\def\tablecellcolor#1{\ifnum #1>100\cellcolor{black!100}\else\cellcolor{black!#1}\fi\ifnum #1>49\color{white}\fi{#1}}
\def\tablecellcolorRQ6#1{\cellcolor{gray!#1}{#1}}

\definecolor{verylightgray}{rgb}{0.93, 0.93, 0.93}

\usepackage[skins]{tcolorbox}

\usepackage{tcolorbox}
\tcbuselibrary{skins}
\newenvironment{summarybox}
{\begin{tcolorbox}
[enhanced,arc=0mm,colback=gray!10,frame hidden,overlay unbroken={%
    \draw[thick,black] (interior.north west)--(interior.south west);
},left=2pt,right=0pt,top=0pt,bottom=0pt,before={\vspace{3pt}\noindent},after={\vspace{0pt}}]}
{\end{tcolorbox}}
\newenvironment{summary}
{\vspace{5pt}\noindent\begin{summarybox}}
{\end{summarybox}\vspace{-5pt}}

\usepackage{multicol}
\usepackage{multirow}
\usepackage{longtable}
\usepackage{lipsum}

\makeatletter
\newcommand\notsotiny{\@setfontsize\notsotiny{6.5}{7}}
\newcommand\almosttiny{\@setfontsize\almosttiny{5.7}{6}}
\makeatother

\newcommand\numberToBeChecked[1]{\textcolor{black}{#1}}

\newcommand\waitingReview[1]{\textcolor{black}{#1}}

\newcommand\datasetLink{https://codeupcrc.github.io}
\newcommand*{\inlineReviewLink}[1]{%
\href{https://codeupcrc.github.io?key=#1}{key=#1}}
\newcommand*{\inlineReviewHref}[2]{%
\href{https://delanohelio.github.io/code_reviews/inlineReviewPage.html?id=#1}{#2}}
\newcommand*{\projectHref}[1]{%
\href{https://github.com/#1}{#1}}

\newcommand*{\tabindent}{ \hspace{3mm}}

\begin{document}

\title{Understanding Code Understandability Improvements in Code Reviews}

\author{Delano Oliveira, Reydne Santos, Benedito de Oliveira, Martin Monperrus, Fernando Castor, and Fernanda Madeiral

\thanks{Delano Oliveira is with the Federal University of Pernambuco, 50732-970 Recife, Brazil, and the Federal Institute of Pernambuco, 55540-000 Palmares, Brazil (e-mail: delanohelio@gmail.com).}
\thanks{Reydne Santos and Benedito de Oliveira are with the Federal University of Pernambuco, 50732-970 Recife, Brazil (e-mail: rbs8@cin.ufpe.br; bfao@cin.ufpe.br).}
\thanks{Martin Monperrus is with the KTH Royal Institute of Technology, 114 28 Stockholm, Sweden (e-mail: monperrus@kth.se).}
\thanks{Fernando Castor is with the University of Twente, 7522 NB Enschede, The Netherlands, and the Federal University of Pernambuco, 50732-970 Recife, Brazil (e-mail: f.castor@utwente.nl).}
\thanks{Fernanda Madeiral is with the Vrije Universiteit Amsterdam, 1081 HV Amsterdam, The Netherlands
(e-mail: fer.madeiral@gmail.com).}
\thanks{Manuscript received 28 August 2023; revised 8 August 2024; accepted 16
August 2024. Recommended for acceptance by K. Blincoe}
\thanks{(Corresponding author: Delano Oliveira.)}
\thanks{Digital Object Identifier no. 10.1109/TSE.2024.3453783}
}

\IEEEtitleabstractindextext{
\begin{abstract}
\textit{Context:} Code understandability plays a crucial role in software development, as developers spend between 58\% and 70\% of their time reading source code. Improving code understandability can lead to enhanced productivity and save maintenance costs.
\textit{Problem:} Experimental studies aim to establish what makes code more or less understandable in a controlled setting, but ignore that what makes code easier to understand in the real world also depends on extraneous elements such as developers' background and project culture and guidelines. Not accounting for the influence of these factors may lead to results that are sound but have little external validity.
\textit{Goal:} We aim to investigate how developers improve code understandability during software development through code review comments. Our assumption is that code reviewers are specialists in code quality within a project.
\textit{Method and Results:} We manually analyzed \numberToBeChecked{2,401} code review comments from Java open-source projects on GitHub and found that over \numberToBeChecked{42\%} of all comments focus on improving code understandability, demonstrating the significance of this quality attribute in code reviews. We further explored a subset of \numberToBeChecked{385} comments related to code understandability and identified \numberToBeChecked{eight} categories of code understandability concerns, such as incomplete or inadequate code documentation, bad identifier, and unnecessary code. Among the suggestions to improve code understandability, \numberToBeChecked{83.9\%} were accepted and integrated into the codebase. Among these, only \numberToBeChecked{two (less than 1\%)} ended up being reverted. We also identified types of patches that improve code understandability, ranging from simple changes (e.g., removing unused code) to more context-dependent improvements (e.g., replacing method calling chains by existing API). Finally, we investigated the potential coverage of four well-known linters to flag the identified code understandability issues. These linters cover less than \numberToBeChecked{30\%} of these issues, although some of them could be easily added as new rules.
\textit{Implications:} Our findings motivate and provide practical insight for the construction of tools to make code more understandable, e.g., understandability improvements are rarely reverted and thus can be used as reliable training data for specialized ML-based tools. This is also supported by our dataset, which can be used to train such models. Finally, our findings can also serve as a basis to develop evidence-based code style guides.
\textit{Data Availability:} Our data is publicly available at \url{\datasetLink}.
\end{abstract}

}

\maketitle

\IEEEdisplaynontitleabstractindextext

\IEEEpeerreviewmaketitle

\vspace*{-50pt}

\section{Introduction}

\IEEEPARstart{R}{eading} code is the main activity of software developers~\cite{siegmund2016}, with  studies~\cite{minelli2015,xia2017} estimating that developers spend between 58\% and 70\% of their time understanding source code. That means that easy-to-read code can save time and money in software development. Code understandability is the extent to which a developer can comprehend and make sense of source code. It covers a number of aspects, such as naming, commenting, and structuring~\cite{Oliveira2020}.
Empirical studies, e.g., \cite{dosSantos2018, Fakhoury2019b, Medeiros2019, Wiese2019, langhout2021}, have compared several different ways of writing code to find which factors impact code understandability, e.g., abbreviated vs. complete word identifier names~\cite{Scanniello2013}.

Laboratory studies on code understandability have as their main advantage the possibility to account for and control extraneous factors and objectively establish relationships between independent and response variables. 
However, they are not able to account for elements such as project culture and guidelines, and developers' background, which directly impact how developers understand code. Not accounting for the influence of these factors may lead to results that are sound but have little external validity.
In this paper, our motivation is to unveil what developers care about regarding code understandability and how they improve it, in a real-world setting. Our key insight is that this knowledge is available through publicly available code reviews in open-source software, where code reviewers act as project specialists in code quality.

Code review is a process in software development where developers read and comment on code written by fellow developers in the same project~\cite{Bacchelli2013,Sadowski2018,Cunha2021}.
In a typical code review scenario, code changes are proposed through pull requests (e.g., in projects hosted on GitHub) and then reviewed by developers who suggest improvements. Once the concerns are addressed and the improvements are performed, the pull request is approved for integration into the codebase. These suggestions might concern several code quality aspects, including understandability.
In this paper, we examine code understandability improvements grounded on real comments about code quality made during code review. More specifically, we investigate (i) what code understandability issues are reported by developers during code review, and (ii) what changes developers apply to address those concerns.

Our methodology is founded on systematic manual analysis of comments written by developers during code review in OSS repositories.
We selected \numberToBeChecked{363} Java open-source projects from GitHub that have an active code review process. In the context of pull requests on GitHub, developers can write general or inline comments. The latter are associated with specific lines of the source code under review. Because we want to identify where the reviewers' concerns reside and what source code changes were applied to address these concerns, we exclusively focused on \textit{inline code review comments} in this study\footnote{Hereafter, we use the term code review comment for simplification.}. We selected a representative sample of \numberToBeChecked{2,401} code review comments for our study.
We first manually analyzed all those code review comments and identified the ones concerning code understandability. From this process, we identified \numberToBeChecked{1,012} code review comments related to code understandability improvements. Then, we analyzed these comments and the corresponding remediation changes that developers made to improve code understandability.

We found that more than \numberToBeChecked{42\%} of all analyzed code review comments suggest code understandability improvements in scopes such as executable source code, code comments, and text literals. This finding highlights the significance of improving code understandability during code review, as suggested by Bacchelli and Bird~\cite{Bacchelli2013} and Sadowski et al.~\cite{Sadowski2018}.
We also analyzed in more depth a random sample of \numberToBeChecked{385} code review comments suggesting code understandability improvements. We identified and labeled the concerns related to code understandability, hereafter referred to as ``understandability smells'', that were mentioned by reviewers in our sample of \numberToBeChecked{385} code review comments. 
We discover \numberToBeChecked{eight} categories of code understandability smells (e.g., \textsc{incomplete or inadequate code documentation}, \textsc{bad identifier}, \textsc{unnecessary code}) in \numberToBeChecked{16} places in the source code (e.g., Method, Call, Literal, Class). Out of these code review comments, \numberToBeChecked{3.9\%} (\numberToBeChecked{323}) of the suggestions made by the reviewers were accepted, indicating that developers generally agree with suggestions for improving code understandability. 
In addition, we performed a fine-grained analysis of actual improvements (the \numberToBeChecked{323} accepted code review comments) and identified different types of patches that improve code understandability (e.g., modify an identifier to express the meaning or type of an element,  remove non-referred constants and other code elements,  add or remove horizontal spacing). While some of these patches can be automatically detected and fixed by linters (e.g., unused variables), others are more challenging to identify as they relate to natural language components (e.g., incomplete code documentation) or are dependent on the context of the project and code style choices (e.g., using boolean expressions instead of equivalent functional methods). 
Furthermore, \numberToBeChecked{97.2\%} of the applied patches for code review comments were integrated into the codebase of the project and they are very rarely reverted (less than 1\%). 
The remaining \numberToBeChecked{2.8\%} of patches had their code removed or replaced with a different solution.
Finally, an examination of four popular linters shows that they can detect less than \numberToBeChecked{30\%} of the analyzed code understandability issues. Some additional cases could be supported by the definition of new rules, whereas others would require them to also interact with natural language text (e.g., to detect typos).
Our study provides practical guidance for the construction of tools to make code more understandable, e.g., some improvements that could be automated are currently not available in any of four popular linters, and understandability improvements are rarely reverted and thus can be used as reliable training data for specialized ML-based tools. This is also supported by our dataset, which can be used to train such models. Finally, our findings can serve as a basis to develop evidence-based code style guides. More broadly, this is a step in helping developers focus on more critical tasks during code review, such as verifying the correctness of code or identifying security vulnerabilities.

\vspace{5pt}
\noindent\textbf{Data Availability.} The resulting dataset of inline code review comments, which contains a classification of developers' intention with a focus on code understandability improvements and the resulting patches created to address them, is publicly available at \url{\datasetLink}.

\section{Background}

This section provides a brief overview of the two concepts that underpin our investigation: code review and code understandability.

\subsection{Code Review}

\begin{figure}[t]
  \centering
  \fbox{\includegraphics[width=0.75\linewidth]{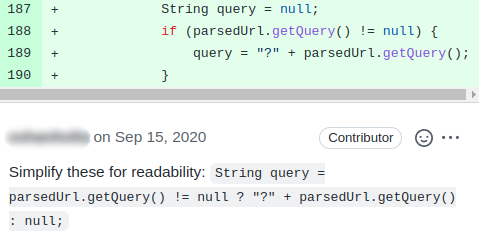}}
  \caption{Inline code review suggestion to replace an \texttt{if} statement by a ternary expression~\cite{r409002709}.}
  \label{fig:example_to_ternary}
  \vspace{-13pt}
\end{figure}

Code review is an assessment of source code to identify a variety of quality issues, such as defects, security vulnerabilities, poor understandability, and bad design.  
In collaborative development platforms, code review takes place when source code changes are proposed to a project by a developer, the \textit{submitter}. Then, the proposed changes are reviewed by other developers, the \textit{reviewers}, who provide comments with, for instance, suggestions or requests for improvements that the submitter should apply to the code.
Submitters and reviewers interact until the reviewers consider that the proposed code changes are good enough to be integrated into the codebase (else the proposed changes are rejected).

GitHub, the most popular collaborative software development platform\footnote{\url{https://github.blog/2023-01-25-100-million-developers-and-counting/}}, allows code review in the context of pull requests. Pull request reviewers can provide general code review comments in the body of the pull request timeline or \textit{inline code review comments} that are specific to source code lines of the changed files.
Inline comments are about fine-grained, localized reviewers' concerns, because they explicitly point to specific source code lines where the concerns lie.
\autoref{fig:example_to_ternary} shows an example of an inline code review comment. The source code lines highlighted in green represent a change proposed to a GitHub project through a pull request. A reviewer wrote an inline comment associated with line 190, suggesting the replacement of the \texttt{if} block by a ternary expression. In such a case, the reviewer's concern is explicitly about readability.


In this paper, we leverage code review comments to investigate the reviewers' concerns related to code understandability and how the submitters address these concerns. Because we want to examine the solutions developers adopt at a fine level of granularity, e.g., the replacement of an \texttt{if} statement by a ternary expression (\autoref{fig:example_to_ternary}), we focus on inline code review comments (we use the term code review comment for simplification). Since code reviewers are quality gatekeepers in software development, we hypothesize that, if code understandability is an important concern in practice, there are code review comments about it.

\subsection{Code Understandability}\label{sec:code_understandability}

Legibility and readability are both important factors for making code easy to read, but understandability is the ultimate goal. Legibility refers to how easily individual code elements can be distinguished from each other, while readability refers to how easily developers can comprehend the code as a whole~\cite{Oliveira2020}. Ultimately, understandability encompasses both legibility and readability and refers to how easily developers can extract relevant information from the source code, including non-functional parts, such as documentation, to perform software development or maintenance tasks~\cite{scalabrino2019}. In this work, we are interested in understandability.

In the literature, many studies have evaluated how to improve code legibility and readability~\cite{Woodfield1981,Tenny1988,Wiese2019,Medeiros2019,Gopstein2017,Hofmeister2019}. All these studies are based on an informal notion of \textit{understandability improvement}. 
A code understandability improvement is a refactoring that, when applied to a program, makes its source code better at communicating to the developer what it does, according to one of more quality criteria~\cite{Oliveira2020}.

A basic example of this type of change is renaming. Renaming a program element, in general, is not a functional correction and does not change the functionality of the program. However, a change to the name of a program element may make it clearer for other humans, by better communicating purpose or rationale. The main motivation for renaming is to improve code understandability, although there are some exceptions where renaming is not motivated by understandability, such as when it matches a reflection pattern. Moreover, certain aspects of source code are pertinent to understandability by nature, specifically documentation and formatting. Although disregarded by compilers, these elements are deliberately integrated into the source code to assist developers in comprehending the code.

Many factors in the source code can affect code understandability, such as the number of identifiers, size of a line of code, and nesting \cite{scalabrino2017, buse2010}. However, the definition of whether a change is an understandability improvement depends on the developer's background (e.g., skill and experience)~\cite{scalabrino2019}. Also, we could consider this can be affected by the developer's goals (e.g., implementing a new feature, writing tests, and fixing a bug).
In the literature, for example, Gopstein et al.~\cite{Gopstein2017} found that subjects misunderstand the output of a code snippet when it contains ternary/conditional expression (e.g., \texttt{a?b:c}) instead of an \texttt{if} statement. In a similar study, Langhout and Aniche~\cite{langhout2021} did not find that the ternary operator causes misunderstanding. In a study by Medeiros et al.~\cite{Medeiros2019}, the practitioners answered that ternary/conditional expression does not influence code understandability positively or negatively.

\begin{figure}[t]
  \centering
  \fbox{\includegraphics[width=0.8\linewidth]{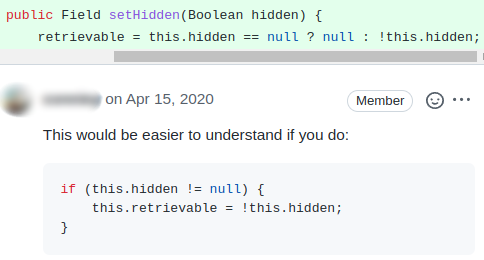}}
  \caption{Inline code review suggestion to replace a ternary expression by an \texttt{if} statement~\cite{r489083957}.}
  \label{fig:example_from_ternary}
  \vspace{-13pt}
\end{figure}

The perception of what understandability improvement is depends on the context in the real world. During code review, reviewers suggest source code changes to improve code understandability, but the development team can influence what is considered an improvement. For instance, in \autoref{fig:example_to_ternary}, a reviewer suggested replacing an \texttt{if} statement by a ternary expression to improve code understandability. In \autoref{fig:example_from_ternary}, which refers to another open-source project, a reviewer suggested replacing a ternary expression by an \texttt{if} statement to improve understandability.
These two code reviewers suggest the opposite operations, both concerning understandability. This means they perceive understandability improvements (in the context of the usage of \texttt{if} statements vs. ternary expression) in a different way.

Style guides (a.k.a. code conventions) also are adopted to improve code understandability in the real world. Smith et al.~\cite{smit2011} defined that \textit{``code conventions are a body of advice on lexical and syntactic aspects of code, aiming to standardize low-level code design under the assumption that such a systematic approach will make code easier to read, understand, and maintain''}. However, the recommendations proposed by different style guides are not only about understandability and can also be divergent. For example, while the Google Java Style Guide\footnote{\url{https://google.github.io/styleguide/javaguide.html}} suggests the use of braces even when the block is empty or contains only a single statement (e.g., \texttt{if (a) \{b+c\}}), the Linux Kernel Coding Style\footnote{\url{https://www.kernel.org/doc/html/v4.10/process/coding-style.html}} disagrees with it for single statements (e.g., \texttt{if (a) b+c}). This suggests that some code alternatives that make source code more understandable are, to some extent, in the eye of the beholder, e.g., they depend on the experience and context of the developer. However, there is insufficient empirical evidence of which code alternatives are preferred during software development considering code understandability to support the creation of a style guide. In this study, we investigate those code alternatives to further help developers choose the best practices and tool builders to create automated support based on empirical evidence from software development teams. We do so in the context of code review, where developers are quality gatekeepers who review and discuss changes until they consider the code is good to be merged into their projects.

\section{Study design}\label{study-design}

This section presents the research questions we aim to answer~(\autoref{sec:rqs}), the data collected for the study (\autoref{sec:projects}), and the process for identifying code review comments that suggest understandability improvements (\autoref{sec:indf_understandability_impro}).

\subsection{Research Questions}\label{sec:rqs}

\newcommand{\rqone}{How often do reviewers ask for code understandability improvements in code review comments?\xspace}

\newcommand{\rqtwo}{What are the main issues pertaining to code understandability in code review?\xspace}

\newcommand{\rqthree}{How likely are understandability improvement comments to be accepted?\xspace}

\newcommand{\rqfour}{What code changes are found in understandability improvements?\xspace}

\newcommand{\rqfive}{To what extent are accepted code understandability improvements reverted?\xspace}

\newcommand{\rqsix}{Do linters contain rules to detect the identified code understandability smells?\xspace}

\vspace{5pt}
\noindent\textbf{RQ1.} \textit{\rqone}

\vspace{3pt}
\noindent\textit{Motivation:} While code reviews are commonly employed to enhance code quality~\cite{Bacchelli2013,Sadowski2018}, there is a lack of information regarding the prevalence of code review comments focusing on code understandability. If a substantial portion of code review comments involves understandability improvements, it may be advantageous to prioritize automation strategies tailored to address this specific area. Furthermore, not every improvement targets executable source code and it may be useful to quantify the extent to which other scopes, such as code comments and string literals, are targeted. Such targeted approaches could lead to significant time and effort savings.

\vspace{3pt}
\noindent\textit{Method:} We analyzed a representative sample of code review comments to determine the rate at which reviewers suggest code understandability improvements. In this direction, we manually classified 2,401 code review comments selected as explained in \autoref{sec:projects}, assessing whether each comment pertains to improving code understandability. Furthermore, we categorized the associated scopes of these comments (e.g., documentation, executable source code, string literals). 
Each code review comment was classified by two authors of this paper. We calculated the Kappa coefficient \cite{Cohen1960} to assess the agreement between their evaluation. We found $k = 0.59$, which is considered a moderate agreement strength \cite{Kitchenham2015}. For every case where there was a disagreement, three or more authors engaged in discussion to reach a final classification. Besides this classification, we also verify the scope to which each comment refers, i.e., where in the source file the reviewer has identified an opportunity to improve understandability (e.g., source code and Javadoc).
Additionally, we randomly selected and classified \numberToBeChecked{385} code review comments related to code understandability. We distinguished between explicit comments, which indicate that the improvement suggestion is motivated by code understandability, and implicit comments, where the intention of code understandability improvement is inferred from the comment text, code snippet, and the code improvement itself. We also provided a description of why we considered each code review comment to be related to code understandability. In the sample, we found a total of \numberToBeChecked{50} explicit comments and \numberToBeChecked{335} implicit comments related to code understandability.

\vspace{3pt}
\noindent\textit{Novelty:} Previous studies~\cite{piantadosi2020,Fakhoury2019,roy2020} explored the projects' files histories to identify code understandability improvements based on state-of-the-art metrics or commit messages that reported an understandability improvement explicitly. We evaluate \numberToBeChecked{2,401} code review comments and identify suggestions by code quality specialists (reviewers) to improve source code understandability. Unlike previous work~\cite{Dantas2023}, we also examine when the intention to improve code understandability is implicit.

\vspace{10pt}
\noindent\textbf{RQ2.} \textit{\rqtwo}    

\vspace{3pt}
\noindent\textit{Motivation:} Identifying the typical understandability issues reported by reviewers can guide researchers and tool builders when designing and training machine learning-based tools to improve understandability. Additionally, this knowledge can contribute to the development of code review checklists~\cite{Chong:2021:ASU} to make the process more systematic and efficient to address code understandability concerns in a structured manner~\cite{Gawande:2011:CMH}.

\vspace{3pt}
\noindent\textit{Method:} 
We selected \numberToBeChecked{385} code review comments randomly from the \numberToBeChecked{1,012} reviews that were related to code understandability improvements and were classified in \autoref{results:rq1}. This sample size of code review comments is a representative sample with a confidence level of \numberToBeChecked{95\%} and a margin of error of \numberToBeChecked{5\%} of the unlimited population, according to Cochran's formula~\cite{israel1992determining}. We then identified the issues that triggered the suggestions for change (the what) and grouped them into categories that we called \textit{understandability smells}. For example, in \inlineReviewHref{MDI0OlB1bGxSZXF1ZXN0UmV2aWV3Q29tbWVudDQzMzc3MjkzOA==}{one code review} [\inlineReviewLink{821}], the reviewer asked to remove three code elements because they were constants that were not being used, i.e., \textsc{unnecessary code} (the ``what''). Also, we identified the type of code element targeted for improvement (the ``where'') using Spoon~\cite{spoon}, a well-established open-source library for analyzing Java source code. Three authors examined the sample of \numberToBeChecked{385} code review comments. They described the issue presented in each comment and identified where the issue occurred in the code. The same authors then discussed the issues and grouped similar comments together. Afterwards, the groups were refined, and categories were established during meetings involving at least five authors.

We used the nodes of Spoon's AST to group constructs related to each issue pointed out by code reviewers into categories. For example, we group the nodes \texttt{CtIf} and \texttt{CtSwitch} into the Conditional category\footnote{The \texttt{CtIf} node refers to an '\texttt{if}' statement in the AST, while the \texttt{CtSwitch} node refers to a '\texttt{switch}' statement. Both are conditional statements in the Java language.}. 
We considered the most specific AST node to which an understandability smell is associated to determine where it happens. For example, in \inlineReviewHref{MDI0OlB1bGxSZXF1ZXN0UmV2aWV3Q29tbWVudDM3NTAyMjI2Nw==}{a code review} [\inlineReviewLink{123}], a reviewer requested using an alternative method invocation to evaluate an expression in an \texttt{if} statement, and we categorize the most specific node identified by Spoon, \texttt{CtInvocation}, as a Call. \waitingReview{However, in other scenarios, the \numberToBeChecked{understandability smell} consists of a set of nodes, and we considered the node that is the closest parent of all other nodes. For instance, \inlineReviewHref{MDI0OlB1bGxSZXF1ZXN0UmV2aWV3Q29tbWVudDQ0MTQ3NzA2NQ==}{in another code review} [\inlineReviewLink{1878}], the reviewer asked to rewrite part of the method's body, and the AST was composed by several nodes such as an invocation and an \texttt{if} statement. In this case, the closest parent of those nodes is a Method, represented by \texttt{CtMethod} in Spoon.} We also categorized understandability smells as Method when the reviewer requested renaming the method, adding Javadoc to it, or moving it. In two code review comments where the reviewers pointed out multiple issues in different parts of the source code, \waitingReview{we classified the ``what'' and ``where'' exclusively based on the location where the code review comment was marked}. For example, in a code review comment in the class declaration where the reviewer asked to \inlineReviewHref{MDI0OlB1bGxSZXF1ZXN0UmV2aWV3Q29tbWVudDQ1MDgyMTc3NA==}{\textit{``remove the Db [from class name] and name attributes from the annotation''}} [\inlineReviewLink{1772}], the code review comment is at the line of the class declaration and we considered that this (Class) is the place where the issue was pointed out.

\vspace{3pt}
\noindent\textit{Novelty:} Previous studies~\cite{piantadosi2020,Fakhoury2019,roy2020} classified the issues related to code understandability in terms of software maintenance types and rules of static analysis tools. However, their classification is limited to these terms, which do not capture the actual code understandability issues. In this work we examined code review comments, as in the work of Dantas et al.~\cite{Dantas2023}, but take two complementary perspectives: coarser-grained, by eliciting the high-level issues that trigger improvements, and finer-grained, by examining which source code elements are connected to these issues. 

\vspace{10pt}
\noindent\textbf{RQ3.} \textit{\rqthree} 

\vspace{3pt}
\noindent\textit{Motivation:} Analyzing the odds of developers accepting comments when they suggest understandability improvements highlights the importance or lack thereof of this kind of comment. 
\waitingReview{It might provide evidence of cases where new tools could be created or not for automatically addressing the identified problems.}

\vspace{3pt}
\noindent\textit{Method:} We aimed to investigate if recommendations for improving code understandability are more likely to be accepted than code reviews that do not address code understandability. To achieve this, we took the pool of \numberToBeChecked{385} code reviews that were analyzed for RQ2 and added another random sample of \numberToBeChecked{385} code reviews that suggest changes unrelated to code understandability.
The sample was divided among the three first authors. \waitingReview{In the first step, the authors searched for revisions in the part of the code where the reviewer left their comments, looking at each new version of the file in the pull request. Eventually, it was necessary to look at other parts of the code that were indicated by the review comments. In the second step, the authors determined whether the reviewer comments were addressed (accepted) by the developers in any of those revisions.} We classified improvement suggestions in reviews as either accepted or not accepted based on whether or not we find the code changes as suggested by the reviewer. We calculated the odds ratio to determine the likelihood of accepting a suggestion, given it is a code understandability improvement.

\vspace{3pt}
\noindent\textit{Novelty:} Brown and Parnin~\cite{brown2020} investigated the acceptability of \textit{GitHub suggested changes}, a special kind of code review comment where reviewers recommend a specific change to the code line or lines. However, as far as we know, no study evaluated how developers addressed reviewer's suggestions for code understandability improvements.

\vspace{10pt}
\noindent\textbf{RQ4.} \textit{\rqfour}

\vspace{3pt}
\noindent\textit{Motivation:} There are multiple studies~\cite{Woodfield1981,Tenny1988,Wiese2019,Medeiros2019,Gopstein2017,Hofmeister2019,Torres:2023:ICC} that empirically evaluated the impact of different ways of writing code on understandability in a lab setting. However, understandability strongly depends on the context where the code must be understood, e.g., due to the experience of the readers. This RQ aims to help developers and researchers to know what makes the code more understandable in practice. Information about typical real-world understandability improvements can help us establish what kinds of tools can address each scenario (linters, more in-depth static analysis, LLM-based, etc.). 

\vspace{3pt}
\noindent\textit{Method:} 
We took as a starting point the set of code reviews examined for RQ2 and RQ3. For this question, we focused on the code review comments where the recommendation made by the reviewer was accepted, and a patch was applied as a consequence. We analyzed the code change applied in each patch to resolve the problem raised by the reviewer. The sample was divided among the three primary authors, who described the improvements made in response to the reviewer's comments. We then organized these different types of changes into categories based on the understandability smells identified in \numberToBeChecked{RQ2}. Finally, we discussed how the improvements could be organized in meetings with at least four of the authors.

\vspace{3pt}
\noindent\textit{Novelty:} As with \numberToBeChecked{RQ2}, we adopted a bottom-up approach where we identify code understandability improvements in terms of the fine-grained refactorings that developers perform. Previous studies~\cite{piantadosi2020,Fakhoury2019} employed a top-down approach, classifying the understandability smells in terms of software maintenance types and rules of static analysis tools. 

\vspace{10pt}
\noindent\textbf{RQ5.} \textit{\rqfive}

\vspace{3pt}
\noindent\textit{Motivation:} Understanding the frequency at which code understandability improvement patches are reverted after being applied during the code review process is essential for assessing the subjectivity of code understandability improvements. Frequent reversions of such patches may indicate a high level of ambiguity and variability in the perception of code understandability, which can lead to challenges when utilizing this data to train machine learning systems, potentially resulting in suboptimal performance~\cite{Nucci:2018:DCS}.

\vspace{3pt}
\noindent\textit{Method:}
For this RQ, we employed the same set of code review comments that were analyzed for RQ4, i.e., comments about understandability where the recommendation made by the reviewer was accepted and a patch was applied. We manually searched subsequent versions of the files modified in that patch to determine whether the patch had been reverted. We considered that a reversion happens when (i) a reviewer suggests improving code understandability, (ii) that suggestion is accepted, and a patch is submitted as a consequence, incorporating said suggestion, and (iii) subsequently, a new patch is submitted, undoing the understandability improvement. In this process, we account for renamed files and cases where portions that were affected by the patch were moved to other parts of the system, e.g., a different file. If the latest version of the file still includes the patch that was introduced as a response to the suggestion for improvement, we considered that it was not reverted. If the file as a whole or the element (method, class, etc.) where the modification was performed is deleted, we also considered that the change was not reverted since we cannot ascertain whether the deletion is connected to said change.

\vspace{3pt}
\noindent\textit{Novelty:} Kalliamvakou et al.~\cite{Kalliamvakou2014} explored the integration rate of proposed changes in GitHub pull requests into the codebase, while Shimagaki et al.~\cite{shimagaki2016} assessed the frequency of commits that were subsequently reverted in project histories. However, a gap remains in the existing literature, as no study specifically examines the adoption and reversion of code understandability improvements within the context of code reviews and the history of the codebase.

\vspace{10pt}
\noindent\textbf{RQ6.} \textit{\rqsix}

\vspace{3pt}
\noindent\textit{Motivation:} Linters are widely used in real-world software development~\cite{Sadowski:2015:TBP,vassallo2020,lenarduzzi2023}. Furthermore, they are capable of detecting several typical problems in source code. It is then worth investigating the extent to which they capture the code understandability issues that code reviewers point out. Ideally, reviewers should not have to worry about problems that an automated tool can capture.  

\vspace{3pt}
\noindent\textit{Method:} We investigated the coverage of the \numberToBeChecked{code understandability smells} identified in our study by rules implemented in Spotbugs~\cite{spotbugs}, PMD~\cite{pmd}, SonarQube~\cite{sonarqube}, and Checkstyle~\cite{checkstyle}. These \numberToBeChecked{four} tools have the highest number of rules~\cite{lenarduzzi2023}, are widely used by practitioners~\cite{vassallo2020}, and have been studied by researchers~\cite{stefanovic2020}. Together, they have a total of \numberToBeChecked{1,315} rules. We manually checked each observation of an \numberToBeChecked{understandability smell} against each tool to determine if it could detect it.

We utilized the set of code review comments that were previously analyzed for RQ4, which are about understandability smell instances. For each understandability smell, we manually inspected the descriptions of the rules implemented in linters to find one that could detect the smell. In the end, we calculated the coverage of understandability smells by computing the number of instances that could be detected by at least one of the linters.

\vspace{3pt}
\noindent\textit{Novelty:} Previous work~\cite{Dantas2023} investigated the extent to which SonarQube can detect code understandability issues in code reviews by running this tool. In contrast, we manually analyzed the applicability of every rule implemented by four different linters. The issues reported by linters depend on how they are configured~\cite{Tomasdottir:2020:AJL}, as linters sometimes have contradicting rules, e.g., SonarQube includes a rule \textit{Close curly brace and the next "else" [..] should be on two different lines} but also one that ends with \textit{...on the same line}. A manual analysis can capture these inconsistencies. 

\subsection{Data Collection}\label{sec:projects}


To answer our research questions, we selected open-source projects with a code review process active in GitHub, i.e., projects where new pull requests are created frequently. Our focus is on the code review comments made by project contributors regarding the code changes submitted through these pull requests. Furthermore, those projects must be non-trivial software systems, e.g., applications, infrastructure components, libraries, frameworks, with many contributors.

We started out by using SEART~\cite{Dabic2021} to search for project candidates because it offers appropriate search criteria. In SEART, we considered the following criteria: (i) Java projects, (ii) at least ten stars (fixed by SEART), (iii) at least one pull request, (iv) at least ten contributors, and (v) not a fork repository.
We selected projects with ten or more contributors because we need a number of contributors that justifies the code review process. These criteria contribute to filtering out personal projects~\cite{Kalliamvakou2014}. 
SEART returned a list of 9,735 projects.


Next, we filtered the projects to select only those where code changes submitted by developers are regularly reviewed by other contributors. Thus, we selected projects with at least one pull request per month in 2020 that include one or more inline code review comments. These requirements ensure that code reviews are commonplace and happen at the level of source code lines. 
Given that our search was conducted in October 2021, we focused on pull requests submitted during the year 2020. With this filter, we kept 389 projects. 

We further refined this set by removing projects where we can not identify the pull requests that were merged since merging a pull request is a clear criterion to determine whether said pull request was accepted. Merging can be done manually through \textit{git} commands on the local repository and then pushing to the remote. Alternatively, it can be done by using the merge button directly on the GitHub platform. Pull requests with automated merging have a clear indication of this event on the pull request's timeline, with its status updated to "merged." However, identifying projects with manual merging can be challenging as it requires the use of heuristics to determine the commit that merged the pull request, as explained by Kalliamvakou et al.~\cite{Kalliamvakou2014}. We identified \numberToBeChecked{eight} projects (e.g., \projectHref{apache/gobblin} and \projectHref{openjdk/skara}) where the pull requests are not merged automatically and decided to remove them. After performing these filtering steps, we kept \numberToBeChecked{381} projects with an active code review process that integrates their code changes in their main branch through GitHub. 

\def\avgNumberOfCharactersByComment{120.2}
\def\maxNumberOfCharactersByComment{44,043}
\def\rateOfEmptyCommentOnlyWithImages{89.6\%}
\def\rateOfEmptyCommentOnlyWithBulletPoint{4.2\%}
\def\rateOfEmptyCommentOnlyWithBulletEmoji{2.1\%}
\def\rateOfCompletelyEmptyComments{4.2\%}
\def\avgNumberOfCharactersInCommentByDiscussion{224.2}
\def\maxNumberOfCharactersInCommentByDiscussion{45,101}
\def\avgNumberOfCharactersInCodeByDiscussion{53.0}
\def\maxNumberOfCharactersInCodeByDiscussion{22,382}
\def\avgNumberOfCommentsByDiscussion{1.9}
\def\numberOfCommentsInLargestDiscussion{43}

Then, we performed a fine-grained filtering of these \numberToBeChecked{381} projects. We manually analyzed each one and evaluated whether it is mainly composed of code. As a result, we excluded \numberToBeChecked{eight} projects that do not fit our software criteria, such as those classified as documentation. Also, we constrained our investigation to projects where most code review comments are written in English. We removed \numberToBeChecked{seven} projects where most code review comments are not in English. In addition, one project (\projectHref{returntocorp/semgrep}) did not primarily use Java, and another (\projectHref{apollographql/apollo-android}) was transitioning from Java to Kotlin. As a result, we removed both. Finally, the code review comments in one project (\projectHref{fisco-bcos/web3sdk}) are all made by bots, such as \textit{``sonarcloud''}. This project was also removed, leaving us with a final total of 363 projects\footnote{The list of selected projects can be found at \url{\datasetLink/projects.html}}.

\begin{figure}[t]
  \centering
  \fbox{\includegraphics[width=0.75\linewidth]{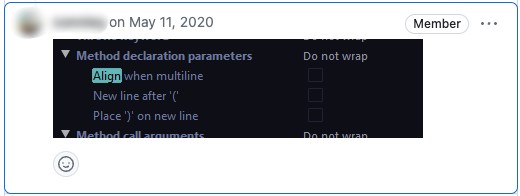}}
  \caption{Example of comment comprised solely of images~\cite{r423093729}.}
  \label{fig:image_only}
  \vspace{-13pt}
\end{figure}

The selected \numberToBeChecked{363} projects comprise more than \numberToBeChecked{349,000} code review comments made in \numberToBeChecked{2020}. 
We downloaded these code review comments using the GitHub GraphQL API\footnote{\url{https://docs.github.com/en/graphql}}. Since we need to manually evaluate and classify the code review comments, we extracted a statistically representative sample. \waitingReview{We chose a confidence level of \numberToBeChecked{95\%} and a margin of error of \numberToBeChecked{2\%}, resulting in a sample size of \numberToBeChecked{2,401} code review comments for an unlimited population size}. \waitingReview{We employed a stratified random sampling~\cite{Groves:2009:SM} strategy, as the resulting sample can be more representative of the investigated population of code review comments than if we just employed simple random sampling~\cite[Chapter~11]{thompson2012sampling}, where projects with few code review comments might not be represented.} 
We randomly selected code review comments from the projects in proportion to the total number of code review comments each project has. As a result, the minimum number of code review comments per project is \numberToBeChecked{1}, the maximum is \numberToBeChecked{15}, and the median is \numberToBeChecked{7}.
The complete list of analyzed comments can be browsed at \url{\datasetLink}. \autoref{tab:descriptive_statistics} presents descriptive statistics of these code review comments and their associated discussions. The table indicates that, in some cases, code review comments can be zero characters. These comments 
comprise only images (\autoref{fig:image_only} presents an example), bullet points without accompanying text, emojis, or are simply empty.

\subsection{Identifying Understandability Improvements}
\label{sec:indf_understandability_impro}

\begin{figure}[t]
    \centering
    \fbox{\includegraphics[width=0.75\linewidth]{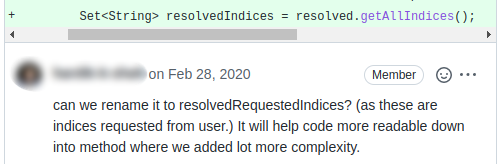}}
    \caption{The reviewer explicitly asks for understandability improvement~\cite{r385993973}.}
    \label{fig:improvement_explicit}
    \vspace{-13pt}
\end{figure}

\begin{figure}[t]
    \centering
    \fbox{\parbox{.75\linewidth}{
    \includegraphics[width=1\linewidth]{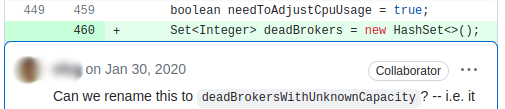}
    
    \includegraphics[width=1\linewidth]{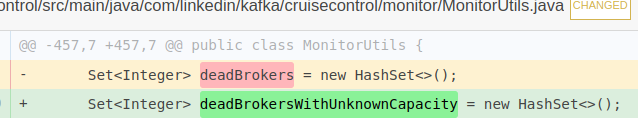}
    }}
    \caption{The reviewer implicitly asks for understandability improvement~\cite{r373266026}.}
    \label{fig:improvement_implicit}
    \vspace{-13pt}
\end{figure}

To answer the research questions of this study, we first classified whether code review comments suggest understandability improvements or not. In this classification process, we need to identify the intention of the reviewer based on their comment. Sometimes, this intention is clear and explicit in the comment text. For example, in \autoref{fig:improvement_explicit}, the reviewer asked to rename a local variable because \textit{``it will help code more readable down into method [...]''}. However, other comments are terse, and the intent is not explicitly stated. For example, in \autoref{fig:improvement_implicit}, the reviewer asks to rename \textit{deadBrokers} to \textit{deadBrokersWithUnknownCapacity} but the intention is not explicit. Although the reviewer does not write it, the most likely motivation is that the new name helps developers better understand the purpose of this variable. 

\begin{table}[tb]
	\caption{\waitingReview{Descriptive statistics for the code review comments, discussions, and source code lines associated to these comments and discussions, for the sample of code reviews in the 363 selected projects.} }\label{tab:descriptive_statistics}
	\centering
	\scriptsize
 \vspace{-5pt}
	\begin{tabular}{@{}l r r r r r r r@{}}
		\toprule
		& Mean & Std & Min & 25\% & 50\% & 75\% & Max \\
		\midrule
            \# comments/project       & 6.6 & 2.3 & 1 & 5 & 7 & 8 & 15 \\
		\# chars/comment       & 120.2 & 216.9 & 0 & 30 & 72 & 148 & 44,043 \\
            \# chars/line of code  & 53.0 & 95.9 & 0 & 30 & 49 & 72 & 22,382 \\
            \# chars/discussion  & 224.2 & 435.7 & 0 & 50 & 115 & 247 & 45,101 \\
            \# comments/discussion & 1.9 & 1.2 & 1 & 1 & 2 & 2 & 43 \\
		\bottomrule
	\end{tabular}
 \vspace{-5pt}
\end{table}

\begin{figure}[t!]
  \centering
  \fbox{\includegraphics[width=0.75\linewidth]{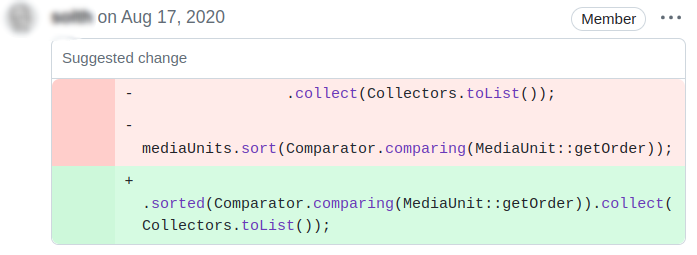}}
  \caption{Example of a suggested change to preserve the function but change the logic~\cite{r471343349}.}
  \label{fig:stream_example}
  \vspace{-13pt}
\end{figure}

\begin{figure}[t!]
    \centering
    \fbox{\parbox{.75\linewidth}{
    \includegraphics[width=1\linewidth]{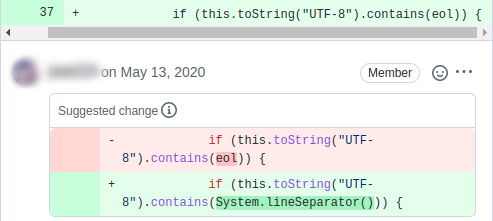}
    
    \includegraphics[width=1\linewidth]{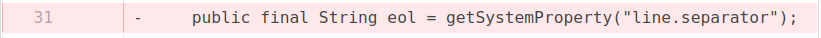}
    }}
    \caption{The intention of the comment is ambiguous~\cite{r424729743}.}
    \label{fig:comment_ambiguous}
    \vspace{-13pt}
\end{figure}

When the inline comment is not enough, it is necessary to consider more information in the context, i.e., the replies to the comment, the source code, and the patch (when available). We first checked whether the suggestion does not change the functionality of the code. For example, in \autoref{fig:comment_ambiguous}, the reviewer asks the author of the pull request to pass another object to the function \texttt{contains} instead of the variable \texttt{eol}. We first evaluated whether the suggestion changes the function of the code snippet to which the suggestion refers because it could indicate that the reviewer is suggesting a correction or different functionality. When we look at the patch, we confirm that the suggestion is behavior-preserving, i.e., the old and the new versions produce the same result given the same input. Thus we consider it as an understandability improvement. However, we found some comments where the logic of the source code changed, but the function is the same. For instance, in \autoref{fig:stream_example}, the reviewer suggests changing how a list is processed in the stream, but the expected result of the stream is preserved. We also consider these cases to be potential understandability improvements.

In summary, we classified a code review comment as suggesting an understandability improvement if it fits within one of the following cases: (i) the intent to improve understandability is explicit in the text, (ii) it is a comment about aspects that do not affect the compilation or the execution of the program, i.e., code comments or code formatting,  (iii) the refactoring that it suggests is motivated mainly by improving understandability, as discussed by Fowler \cite{fowler2018refactoring}, or (iv) it does not explicitly state its intent, but it makes a suggestion that is very similar to a comment where the intent is explicit, i.e., the suggestions for improvement require the same change, e.g., in an explicit code review comment, the type \texttt{Boolean} is replaced by \texttt{boolean} to ensure consistency [\inlineReviewLink{1201}], similarly, in an implicit code review comment, the type \texttt{LinkedList} is replaced by \texttt{ArrayList} [\inlineReviewLink{1089}] for the same reason. 
In a code review comment, the reviewer could suggest an improvement that impacts multiple factors, such as performance and understandability. We also consider those types of improvements that impact understandability.
Each comment, including those where the reviewer explicitly mentioned that it pertains to understandability or those that are naturally about understandability (as described in \autoref{sec:code_understandability}), was classified by at least two authors. Conflicts between answers were resolved through meetings involving at least five authors. \waitingReview{In the meetings, we analyzed the conflicts through discussion sessions and resolutions were achieved by consensus. When consensus could not be achieved, the code review comment was discarded and a new one was selected for labeling.}
In the remainder of this section, we explain the types of comments that required additional analysis.  

\paragraph{Ambiguous intention comments} There are scenarios where it is impossible to discern the intention of the reviewer. For example, in a comment~\cite{r537391027}, the reviewer suggested ``\textit{better to use putString}''. This suggestion can be related to improving code understandability or improving performance. There is no evidence of any intention, neither in replies nor patches. If we cannot establish intention with some confidence, we discard the comment.

\paragraph{Repetition comments} Similarly, in other code review comments, we find tiny comments, e.g., \textit{``ditto''}~\cite{r486429425} and \textit{``same here''}~\cite{r497201087}, that do not bring any information about the reviewer's intention. It would be necessary to look at another code review comment to get the context of the comment, and looking for the original comment could be time-consuming. Since these comments account for approximately 1\% of our sample, we decided to disregard them. However, we kept them in the dataset labeled as ``Discard'' for future research.

\paragraph{Question comments} Sometimes reviewers use the code review comments to ask questions to the PR author. On the one hand, there are clarifying questions where reviewers attempt to understand the purpose of the code snippet under review (e.g., ``\textit{Why is this needed?}''~\cite{r463982271}, ``\textit{Why protected?}''~\cite{r441271747}). In those scenarios, we considered that the comment does not suggest an understandability improvement. On the other hand, we found scenarios where reviewers suggested understandability improvements through questions. For example, in a code review comment, the reviewer asked ``\textit{wrong indent?}''~\cite{r530723377}. This suggestion, phrased as a question, indicates a problem with how the code is formatted. This kind of code review question is considered related to understandability improvements. Previous work~\cite{Ebert:EMSE:2021} has shown that comments with different purposes are often phrased as questions.

\paragraph{Multi-purpose comments} In other code reviews, the reviewers suggested changes with multiple purposes. In one example~\cite{r378437693}, the reviewer commented:   ``\inlineReviewHref{MDI0OlB1bGxSZXF1ZXN0UmV2aWV3Q29tbWVudDM3ODQzNzY5Mw==}{\textit{[...] Suggested change \texttt{if (path.contains("b/")) \{} $\rightarrow$ \texttt{if (path.startsWith("b/"))} \{ for better performance [...] and preferably change the name of the urlString variable back to just url}}'' [\inlineReviewLink{2151}\footnote{Hereafter, we use the reference [key=\#\#\#\#] to represent a link [https://codeupcrc.github.io?key=\#\#\#\#], pointing to the code review comments in our dataset.}]. Although the reviewer suggested a change to improve performance, they also asked to rename an identifier intending to improve the code understandability. This kind of code review comment was considered related to understandability improvements. In our understanding, it is a single comment suggesting multiple different changes. 

\paragraph{Bot comments} Finally, we found code review comments introduced by bots. Some of them are suggestions based on linters to improve understandability. For example, the bot named \textit{codeclimate} created the following code review comment~\cite{r460364607}: ``\href{https://delanohelio.github.io/code_reviews/inlineReviewPagePilot.html?json=https://delanohelio.github.io/code_reviews/triplea-game_triplea/pr_7241.json&inline=MDI0OlB1bGxSZXF1ZXN0UmV2aWV3Q29tbWVudDQ2MDM2NDYwNw==}{\textit{Method buildMenu has 59 lines of code (exceeds 30 allowed). Consider refactoring}}''. Although the suggestion of the code review comment is related to understandability improvement, we discarded these code review comments because there was not an evaluation by a human specialist (i.e., a reviewer). Previous work~\cite{Sadowski:2015:TBP} has shown that many recommendations made by linter bots are ignored by reviewers.

\section{Results}

\subsection{\rqone (RQ1)}\label{results:rq1}

\autoref{tab:rq1:result} presents the results of our manual classification of code review comments in terms of whether they pertain to code understandability or not. Each row indicates the number of comments per scope, which can be related to (i) executable source code (\textsc{Code}), i.e., excluding comments and blank lines and focusing on the instructions that will actually be compiled and executed, (ii) high-level API documentation (\textsc{Javadoc}), (iii) regular source code comments (\textsc{Comment}), (iv) a string value or expression with strings (\textsc{String literal}), or (v) \textsc{Others} (e.g., annotations). We separated the comments to generate documentation and regular comments because they have different characteristics. \textsc{Javadoc} is well structured, has special annotations to reference specific elements in the documentation, and is used to explain classes, attributes, and methods, often for readers who are not looking at the code. \textsc{Comment} has no prescribed structure and is used to explain the code or the rationale behind it in a more direct manner for readers who are directly looking at a specific part of the code. Also, we analyzed \textsc{String literal}s separately because these literals may give developers hints about program behavior, e.g., in log messages, but they are not comments. Finally, annotations, log levels, and other special cases are analyzed separately. We did not consider annotations to be part of the executable source code because, according to the Java Language Specification\footnote{\url{https://docs.oracle.com/javase/specs/jls/se12/html/index.html}}, they have \textit{``no effect at run time''}.

\def\avgCharsUnderstandComment{99.21}
\def\avgCharsNoComment{117.70}

In the analyzed sample, we found out that \numberToBeChecked{1,012} (\numberToBeChecked{42.15\%}) of the code review comments include suggestions, requests, or discussions about code understandability improvements. This amounts to more than \numberToBeChecked{two out of every five} code reviews. This result emphasizes that reviewers are worried not only about the correctness and performance of the code but also whether the developers are following the best practices of programming and code understandability. We check whether comments suggesting code understandability improvements differ from other comments by comparing their sizes. If one kind of comment tends to be significantly shorter, this may suggest that developers are willing to dedicate less time and effort to it. The \numberToBeChecked{1,012} comments suggesting code understandability improvements tend to be shorter (\avgCharsUnderstandComment~characters) than those not related to understandability (\avgCharsNoComment~characters). The difference is statistically significant (Mann-Whitney U, $p=0.000001$) but the effect size is negligible (Cliff's Delta, -0.12). \waitingReview{As for the associated source code lines, i.e., the lines where the inline code reviews were added}, the mean character counts are 50.52 and 54.15 and the difference is not statistically significant (Mann-Whitney U, $p=0.07$). Overall, we can say that the difference between comments and source lines of code associated with understandability improvements and those that are not is inconsequential. 

Considering all code review comments analyzed (2,401), the majority of them (\numberToBeChecked{82.2\%, 1,974 out of 2,401}) are in the scope of \textsc{Code}. Among the \numberToBeChecked{1,974} code review comments whose scope is \textsc{Code},  \numberToBeChecked{33.49\% (661 out of 1,974)} are related to code understandability improvements. The reviewers in these code review comments have asked code submitters to rename identifiers, remove unnecessary code elements, use alternative code constructs, and others. We discuss the content of those code review comments in \autoref{results:rq2}. The results support other studies~\cite{Bacchelli2013,Sadowski2018} that indicate that improving code quality is one of the main goals of code reviews. 

When we look only at the code review comments in the scope of \textsc{Javadoc}, \numberToBeChecked{86.26\% (157 out of 182)} of them are about understandability improvements. Generally, reviewers ask code change submitters to add Javadoc documentation (e.g., \inlineReviewHref{MDI0OlB1bGxSZXF1ZXN0UmV2aWV3Q29tbWVudDUxMjU4NjM4Mw==}{\textit{``Missing javadoc''}} [\inlineReviewLink{747}]), improve it (e.g., \inlineReviewHref{MDI0OlB1bGxSZXF1ZXN0UmV2aWV3Q29tbWVudDQwMzA0MzU2Mg==}{\textit{``@link on this and the next line please''}} [\inlineReviewLink{802}]), or fix the description (e.g., \inlineReviewHref{MDI0OlB1bGxSZXF1ZXN0UmV2aWV3Q29tbWVudDUzODA5ODQ3NA==}{\textit{``appear $\rightarrow$ appears''}} [\inlineReviewLink{819}]). 
We found examples of code review comments related to Javadoc that do not suggest understandability improvements. For example, the code review comment \inlineReviewHref{MDI0OlB1bGxSZXF1ZXN0UmV2aWV3Q29tbWVudDQzNDIyODc2Nw==}{\textit{``what if these objects don't have a common ID or don't have a common event time?''}} [\inlineReviewLink{26}] discusses the design solution described on the Javadoc. In this case, the documentation is not the target of the discussion but the code design related to this documentation.

Similarly to \textsc{Javadoc}, \numberToBeChecked{78.18\% (86 out of 110)} of the code review comments in the scope of \textsc{Comments} are related to understandability improvements. The reviewers suggest the addition (e.g., \inlineReviewHref{MDI0OlB1bGxSZXF1ZXN0UmV2aWV3Q29tbWVudDQxMzk5Nzg5OA==}{\textit{``Please add comments where you see fit.''}} [\inlineReviewLink{1287}]), improvement (e.g., \inlineReviewHref{MDI0OlB1bGxSZXF1ZXN0UmV2aWV3Q29tbWVudDUyNjU3Mzk0MA==}{\textit{``Nit: Typo shake $\rightarrow$ sake''}} [\inlineReviewLink{182}]), or removal (e.g., \inlineReviewHref{MDI0OlB1bGxSZXF1ZXN0UmV2aWV3Q29tbWVudDM3MjE5NDM2Ng==}{\textit{``we should remove this comment''}} [\inlineReviewLink{1062}]) of comments in the source code. Conversely, some of the code review comments that impact comments do not relate to understandability improvements. They are related to comments with tags such as \textit{TODO} and \textit{FIXME}.

\begin{table}[t]
	\caption{Classification of the 2,401 code review comments in terms of whether they pertain to code understandability or not. The two groups are further divided based on the scopes to which the code review comments refer. The percentages for each row add up to 100\%.}\label{tab:rq1:result}
	\centering
	\scriptsize
 \vspace{-5pt}
	\begin{tabular}{@{}l r r r@{}}
		\toprule
                  & \multicolumn{3}{c}{Is it about understandability?} \\
		Scope & \multicolumn{1}{c}{Yes} & \multicolumn{1}{c}{No} & \multicolumn{1}{c}{Discarded} \\
		\midrule
		\textsc{Code} (1,974)         & 33.49\% \hspace*{3.2pt} (661) & 63.93\% (1,262) & 2.58\% (51) \\
            \textsc{Javadoc} (182)       & 86.26\% \hspace*{3.2pt} (157) & 12.09\% \hspace*{6.8pt} (22)    & 1.65\% \hspace*{1.5pt} (3) \\
		\textsc{Comments} (110)      & 78.18\% \hspace*{6.8pt} (86)   & 20.91\% \hspace*{6.8pt} (23)    & 0.91\% \hspace*{1.5pt} (1) \\
		\textsc{String literal} (81) & 80.25\% \hspace*{6.8pt} (65)    & 18.52\% \hspace*{6.8pt} (15)    & 1.23\% \hspace*{1.5pt} (1) \\
		\textsc{Other} (54)          & 79.63\% \hspace*{6.8pt} (43)   & 20.37\% \hspace*{6.8pt} (11)    & 0\% \hspace*{1.5pt} (0) \\
		\midrule
		\textbf{Total (2,401)} & 42.15\% (1,012) & 55.52\% (1,333) & 2.33\% (56) \\
		\bottomrule
	\end{tabular}
 \vspace{-14pt}
\end{table}

In the analyzed code review comments related to \textsc{String literal} scope, \numberToBeChecked{80.25\% (65 out of 81)} can be considered code understandability improvements. In those code reviews, the reviewers suggest changes to string literals to improve exception messages (e.g., \inlineReviewHref{MDI0OlB1bGxSZXF1ZXN0UmV2aWV3Q29tbWVudDM5NTIyODgzNw==}{\textit{``Can you add what the expected and actual partition path was in the exception message ?''}} [\inlineReviewLink{177}]), assertion error messages (e.g., \inlineReviewHref{MDI0OlB1bGxSZXF1ZXN0UmV2aWV3Q29tbWVudDUyMzIzNjY2NA==}{\textit{``Can you please change this to be cookieMaxAge cannot be zero?''}} [\inlineReviewLink{2096}]), logging messages (e.g., \inlineReviewHref{MDI0OlB1bGxSZXF1ZXN0UmV2aWV3Q29tbWVudDU0MzgyMDgwMQ==}{\textit{``I suggest that you can change to \texttt{logger.error("unknown error", e);}''}} [\inlineReviewLink{2294}]), and other method arguments (e.g., \inlineReviewHref{MDI0OlB1bGxSZXF1ZXN0UmV2aWV3Q29tbWVudDM5NzkwMzU0OA==}{\textit{``The number of threads...ingest $\rightarrow$ This property is deprecated since 2.1.0. The number of threads...''}} [\inlineReviewLink{60}]). In contrast, in the code review comments unrelated to understandability improvements, the string literals are arguments used to send messages to the end user (e.g., \inlineReviewHref{MDI0OlB1bGxSZXF1ZXN0UmV2aWV3Q29tbWVudDQ5MzA0NjU5Mw==}{\textit{``Let's change the description to ``List of transports the GAPIC can use...''}} [\inlineReviewLink{1008}]), the paths of a resource on disk (e.g., \inlineReviewHref{MDI0OlB1bGxSZXF1ZXN0UmV2aWV3Q29tbWVudDM2ODQyMTYzMg==}{\textit{``I think you have a typo in the filename: the file is: ClangAnalzer.txt the fileName is: ClangAnalyzer.txt''}} [\inlineReviewLink{1259}]), and tags used on the header of a protocol (e.g., \inlineReviewHref{MDI0OlB1bGxSZXF1ZXN0UmV2aWV3Q29tbWVudDM3NDk2Njc1Mw==}{\textit{``TFB $\rightarrow$ ServiceTalk...''}} in  \texttt{.addHeader("Server", "TFB")} [\inlineReviewLink{2157}]).

In the \textsc{Other} scope, the code review comments are related to annotations and logging (excluding comments that directly refer to string literals). From those, \numberToBeChecked{79.63\% (43 out of 54)} are related to understandability improvements. The reviewers ask code change submitters to change (including adding and removing) logs and their configuration (e.g., \inlineReviewHref{MDI0OlB1bGxSZXF1ZXN0UmV2aWV3Q29tbWVudDQ5MjE2ODQ4MA==}{\textit{``Do we need this log here or was it just for debugging?''}} [\inlineReviewLink{2372}], \inlineReviewHref{MDI0OlB1bGxSZXF1ZXN0UmV2aWV3Q29tbWVudDQ5OTU3NDYwMQ==}{\textit{``Use \texttt{log.info}?''}} [\inlineReviewLink{1896}]) and add documentation annotations (e.g., \inlineReviewHref{MDI0OlB1bGxSZXF1ZXN0UmV2aWV3Q29tbWVudDM4MzI2MzE1Ng==}{\textit{``\texttt{@PublicAPI} instead of \texttt{@beta}''}} [\inlineReviewLink{839}]). We also found code review comments that not support understanding the source code where reviewers ask for modifications in annotations that may be related to bugs (\inlineReviewHref{MDI0OlB1bGxSZXF1ZXN0UmV2aWV3Q29tbWVudDQzNTE5Njk3MA==}{\textit{``Does the lazy inject actually work since \texttt{AuditEventLoggingFacade} is not a lazy bean?''}} [\inlineReviewLink{1545}]) or modifications that are related to release notes (\inlineReviewHref{MDI0OlB1bGxSZXF1ZXN0UmV2aWV3Q29tbWVudDQxODAyNDgwNQ==}{\textit{``An entry in CHANGES\_NEXT\_RELEASE describing the fix...should be included''}} [\inlineReviewLink{2161}]), for example.

\begin{summary}
\begin{small}
\textbf{Answer to RQ1.}
We found that \numberToBeChecked{42.15\% (1,012/2,401)} of the selected discussions are related to code understandability improvements.
Among them, \numberToBeChecked{65.31\% (661/1,012)} are about executable source code, almost \numberToBeChecked{24,01\% (157+86 out of 1,012)} are related to comments, and \numberToBeChecked{more than 10.67\% (65+43 out of 1,012)} are related to literal strings, annotations, and others. Approximately one out of every three code review comments that refer to executable source code is about understandability. Furthermore, \numberToBeChecked{four out of five} code review comments that refer to elements other than executable code in a program source file pertain to understandability improvements.

\vspace{5pt}
\noindent\textbf{Implications.}
In previous research ~\cite{Fakhoury2019,roy2020}, the researchers identified ways to improve code understandability at the level of files by analyzing commit messages. Our findings suggest that researchers and developers can also learn from code reviews to improve code understandability at the level of source code snippets. Moreover, our results demonstrate that automated tools that enhance code understandability can have a significant impact in reducing developers' effort during code reviews. Additionally, our work makes it clear that improving code understandability requires tools that work with both programming language and natural language elements. 
\end{small}
\end{summary}

\subsection{\rqtwo (RQ2)} \label{results:rq2}

\autoref{tab:rq2:result} shows the frequency of the understandability smell (rows) found in our sample and where they are in the source code (columns), with each cell matching the understandability smell in a specific place in the source code. We try to keep the names of the understandability smells as self-explanatory as possible. Therefore, we sometimes employ longish names such as \textit{``Incomplete or inadequate code documentation''}, instead of more typical smell names such as \textit{``Shotgun surgery''} and \textit{``Comments''}~\cite{fowler2018refactoring}. 
We categorized \numberToBeChecked{eight} kinds of \numberToBeChecked{understandability smells} found in \numberToBeChecked{16} places in the source code. More specifically, we encountered the following places (with the number of occurrences in parentheses): Method (\numberToBeChecked{130}), Call (\numberToBeChecked{55}), Literal (\numberToBeChecked{40}), Class (\numberToBeChecked{36}), Attribute (\numberToBeChecked{29}), Conditional (\numberToBeChecked{27}), Variable (\numberToBeChecked{22}), Parameter (\numberToBeChecked{11}), Import (\numberToBeChecked{8}), Operation (\numberToBeChecked{7}), Try-Catch (\numberToBeChecked{7}), Loop (\numberToBeChecked{5}), Annotation (\numberToBeChecked{5}), Interface (\numberToBeChecked{1}), Object (\numberToBeChecked{1}), Package (\numberToBeChecked{1}). The categories with less than \numberToBeChecked{10} instances of understandability smells were grouped under Others in the table. 

The most prevalent \numberToBeChecked{understandability smell} is \textsc{Incomplete or inadequate code documentation}, which appears in  \numberToBeChecked{22.3\% (86 out of 385)} of the code review comments. It groups the cases where the concerns are in documentation and code comments. There are comments where reviewers point out concerns related to \texttt{Javadoc}, inline code comments, and Java annotations. In general, the reviewers indicated missing documentation (e.g., \inlineReviewHref{MDI0OlB1bGxSZXF1ZXN0UmV2aWV3Q29tbWVudDM3NzA2OTAxNA==}{\textit{``Add a comment indicating whether start/end are inclusive or exclusive''}} [\inlineReviewLink{2020}]); incorrect documentation (e.g., \inlineReviewHref{MDI0OlB1bGxSZXF1ZXN0UmV2aWV3Q29tbWVudDM5OTQxOTAxNw==}{\textit{``Minor: which is available...''}} [\inlineReviewLink{1115}]); incorrect placement of documentation (e.g., \inlineReviewHref{MDI0OlB1bGxSZXF1ZXN0UmV2aWV3Q29tbWVudDQwMzc5OTkzOQ==}{\textit{``Think you may have intended to have the below comment here too, or only here since it's the first occurrence''}} [\inlineReviewLink{1724}]); and unnecessary comments (e.g., \inlineReviewHref{MDI0OlB1bGxSZXF1ZXN0UmV2aWV3Q29tbWVudDQwNjQ1NTAzMA==}{\textit{``What does the comment mean? Or can it be removed?''}} [\inlineReviewLink{321}]). Furthermore, the reviewers indicated incorrect or missing documentation about the status of methods using the Java annotation \texttt{@Deprecated}. More than half (\numberToBeChecked{\waitingReview{45/86 = 52.3\%}}) of the occurrences of this \numberToBeChecked{understandability smell} are related to Method (\numberToBeChecked{39} in declaration clause and \numberToBeChecked{six} in body block) and about \numberToBeChecked{17.4\% (15/86)} are related to Class. We find more understandability smells about code documentation in these places because, in Java, code documentation is frequently written in methods and classes.

The second most popular \numberToBeChecked{understandability smell} is \textsc{Bad identifier}, which appears in \numberToBeChecked{\waitingReview{20.3\% (78 out of 385)}} of the code review comments. The reviewers pointed out concerns related to the content of the identifier name and its style. A good identifier can help better communicate the content or its functionality, e.g., the variable \texttt{n} that represents the person's name is better represented by identifier \texttt{name}.  We highlight code review comments 
indicating typos in the identifier names (e.g., \inlineReviewHref{MDI0OlB1bGxSZXF1ZXN0UmV2aWV3Q29tbWVudDQwMTIzNDU4NQ==}{\textit{``Typo in variable name''}} [\inlineReviewLink{811}]); 
parameter names in upper case (e.g., \inlineReviewHref{MDI0OlB1bGxSZXF1ZXN0UmV2aWV3Q29tbWVudDQ0NTgyNzc0OQ==}{\textit{``Would ... making ... the MBI parameter in createPatientResource() lowercase? Might get mistaken for constants...''}} [\inlineReviewLink{579}]); 
method names that do not explain functionality (e.g., \inlineReviewHref{MDI0OlB1bGxSZXF1ZXN0UmV2aWV3Q29tbWVudDUwODIxMzQ4Mw==}{\textit{``This method actually returns the \texttt{Requirements} class, not the annotation, so better call it \texttt{getArtifactRequirements()}''}} [\inlineReviewLink{535}]); 
and variable names incompatible with their type (e.g., \inlineReviewHref{MDI0OlB1bGxSZXF1ZXN0UmV2aWV3Q29tbWVudDUxNzMwNTQxNw==}{\textit{``Rename the field as well? Since it's now an executor and not a handler...''}} [\inlineReviewLink{340}]).
Most of the occurrences of this kind of understandability smell are in Method (\numberToBeChecked{32} in the declaration clause and \numberToBeChecked{two} in the body block), Attribute (\numberToBeChecked{14}), Variable (\numberToBeChecked{11}), and Class (\numberToBeChecked{10}).

Another \numberToBeChecked{understandability smell} is \textsc{complex, long, or inadequate logic}. The reviewers pointed out concerns about complex and long alternatives to writing expressions, code constructs, and other statements. For instance, the boolean expression \texttt{map.keySet().size() > 0} could be replaced by a unique method invocation, \texttt{!map.isEmpty()}. We find this category in \numberToBeChecked{\waitingReview{18.2\% (70/385)}} of the observations in our sample. The reviewers have identified 
redundant method calls (e.g., \inlineReviewHref{MDI0OlB1bGxSZXF1ZXN0UmV2aWV3Q29tbWVudDQ2MzY5NTYwOQ==}{\textit{``Looks like it could be simplified''}} [\inlineReviewLink{83}]); 
and long method implementation (e.g., \inlineReviewHref{MDI0OlB1bGxSZXF1ZXN0UmV2aWV3Q29tbWVudDQ5NDQ0MTY1Mw==}{\textit{``Maybe you can refactor this logic into a different method, it makes this method quite large...''}} [\inlineReviewLink{1413}]). 
We also found comments suggesting that anonymous inner classes are too verbose (e.g., \inlineReviewHref{MDI0OlB1bGxSZXF1ZXN0UmV2aWV3Q29tbWVudDU1NjU4NDEyNw==}{\textit{``With a lambda, this would be slightly shorter''}} [\inlineReviewLink{58}]); 
and complex \texttt{if-else} implementation with repeated code (\inlineReviewHref{MDI0OlB1bGxSZXF1ZXN0UmV2aWV3Q29tbWVudDQ0MTQ3NzA2NQ==}{\textit{``I think a better way to write this method is: [duplicate code within if-else out of if-else statement]''}} [\inlineReviewLink{1878}]). 
Moreover, reviewers indicated concerns related to using API classes, such as using a class as type instead of its primitive type (e.g., \inlineReviewHref{MDI0OlB1bGxSZXF1ZXN0UmV2aWV3Q29tbWVudDUyOTk2OTI5NQ==}{\textit{``Why \texttt{Boolean} and not \texttt{boolean}?''}} [\inlineReviewLink{1201}]). 
This smell is more frequent in Call (\numberToBeChecked{27}), Methods (\numberToBeChecked{eleven} in the body and \numberToBeChecked{two} in the declaration clause), and Conditionals (\numberToBeChecked{nine}).

\begin{table*}[t]
	\caption{The frequency of understandability smells found by reviewers and where they are in the source code. For the code elements marked with an asterisk, the understandability smells were found in their declaration, access, or implementation.}\label{tab:rq2:result}
	\centering
	\scriptsize
 \vspace*{-5pt}
	\begin{tabular}{@{}lccccccccccc@{}}
		\toprule
		Understandability smells & 
            \rotatebox{40}{\parbox{0.8cm}{Method*}} &
            \rotatebox{40}{\parbox{0.8cm}{Call}} & 
            \rotatebox{40}{\parbox{0.8cm}{Literal}} & 
            \rotatebox{40}{\parbox{0.8cm}{Class}} & 
            \rotatebox{40}{\parbox{0.8cm}{Attribute*}} & 
            \rotatebox{40}{\parbox{1.3cm}{Conditional}} & 
            \rotatebox{40}{\parbox{0.8cm}{Variable*}} & 
            \rotatebox{40}{\parbox{0.8cm}{Parameter*}} &
            \rotatebox{40}{\parbox{0.8cm}{Others}} &
            \rotatebox{40}{\parbox{0.8cm}{Total}} \\
		\midrule
	    \textsc{\scriptsize{Incomplete or inadequate code documentation}} &
                \tablecellcolor{45} & \tablecellcolor{4} & \tablecellcolor{1} & \tablecellcolor{15} & \tablecellcolor{3} & \tablecellcolor{9} & \tablecellcolor{3} & \tablecellcolor{0} & \tablecellcolor{6} & \tablecellcolor{86} \\
            \textsc{\scriptsize{Bad identifier}} &
                \tablecellcolor{34} & \tablecellcolor{0} & \tablecellcolor{0} & \tablecellcolor{10} & \tablecellcolor{14} & \tablecellcolor{0} & \tablecellcolor{11} & \tablecellcolor{9} & \tablecellcolor{0} & \tablecellcolor{78} \\
            \textsc{\scriptsize{Complex, long, or inadequate logic}} &
                \tablecellcolor{13} & \tablecellcolor{27} & \tablecellcolor{0} & \tablecellcolor{2} & \tablecellcolor{3} & \tablecellcolor{9} & \tablecellcolor{4} & \tablecellcolor{1} & \tablecellcolor{11} & \tablecellcolor{70} \\
	    \textsc{\scriptsize{Unnecessary Code}} &
                \tablecellcolor{17} & \tablecellcolor{8} & \tablecellcolor{1} & \tablecellcolor{2} & \tablecellcolor{8} & \tablecellcolor{3} & \tablecellcolor{3} & \tablecellcolor{0} & \tablecellcolor{9} & \tablecellcolor{51} \\
            \textsc{\scriptsize{Inconsistent or disrupted formatting}} &
                \tablecellcolor{17} & \tablecellcolor{4} & \tablecellcolor{1} & \tablecellcolor{7} & \tablecellcolor{1} & \tablecellcolor{6} & \tablecellcolor{0} & \tablecellcolor{0} & \tablecellcolor{6} & \tablecellcolor{42} \\
            \textsc{\almosttiny{Wrong, missing, or inadequate string expression or literal}} &
                \tablecellcolor{0} & \tablecellcolor{1} & \tablecellcolor{29} & \tablecellcolor{0} & \tablecellcolor{0} & \tablecellcolor{0} & \tablecellcolor{0} & \tablecellcolor{0} & \tablecellcolor{1} & \tablecellcolor{31} \\
            \textsc{\scriptsize{Inadequate logging and monitoring}} &
                \tablecellcolor{3} & \tablecellcolor{11} & \tablecellcolor{1} & \tablecellcolor{0} & \tablecellcolor{0} & \tablecellcolor{0} & \tablecellcolor{1} & \tablecellcolor{0} & \tablecellcolor{2} & \tablecellcolor{18} \\
            \textsc{\scriptsize{Missing constant usage}} &
                \tablecellcolor{1} & \tablecellcolor{0} & \tablecellcolor{7} & \tablecellcolor{0} & \tablecellcolor{0} & \tablecellcolor{0} & \tablecellcolor{0} & \tablecellcolor{1} & \tablecellcolor{0} & \tablecellcolor{9} \\
        \midrule
            Total &
                \tablecellcolor{130} & \tablecellcolor{55} & \tablecellcolor{40} & \tablecellcolor{36} & \tablecellcolor{29} & \tablecellcolor{27} & \tablecellcolor{22} & \tablecellcolor{11} & \tablecellcolor{35} & \tablecellcolor{385} \\
        \bottomrule
	\end{tabular}
 \vspace*{-13pt}
\end{table*}

\textsc{Unnecessary code} occurs in \numberToBeChecked{\waitingReview{13.2\% (51 out of 385)}} of the code review comments in our sample. This \numberToBeChecked{understandability smell} involves snippets of source code that can be safely removed without altering the functionality of the code. This goes beyond simpler scenarios, such as textually duplicated or dead code. For example, in the code snippet \texttt{V1=2; V1+=1; V1=1}, the two first statements do not influence the final value of \texttt{V1} but may still impose a cognitive load on the reader and, in the worst case, may cause her to misjudge the final value of \texttt{V1}. Among other elements, the reviewers identified 
unused imports, classes, methods, variables, and constants (e.g., \inlineReviewHref{MDI0OlB1bGxSZXF1ZXN0UmV2aWV3Q29tbWVudDU0NzIzNDcwMA==}{\textit{``unused constant''}} [\inlineReviewLink{816}]);  
and commented out source code (e.g., \inlineReviewHref{MDI0OlB1bGxSZXF1ZXN0UmV2aWV3Q29tbWVudDU0Njc4NDIyOQ==}{\textit{``commented code''}} [\inlineReviewLink{1074}]). 
Other concerns are related to redundant code, such as unnecessary use of \texttt{this} (e.g., \inlineReviewHref{MDI0OlB1bGxSZXF1ZXN0UmV2aWV3Q29tbWVudDU1NjcwODAyNw==}{\textit{``nit: this. is redundant here''}} [\inlineReviewLink{1871}]);
unnecessary intermediate variable (e.g., \inlineReviewHref{MDI0OlB1bGxSZXF1ZXN0UmV2aWV3Q29tbWVudDM3MDU2NDEyNQ==}{\textit{``Could be better to have the full \texttt{clientSession.newRequest()...execute()} to avoid having \texttt{loadCsv} variable''}} [\inlineReviewLink{1561}]);
and semantically duplicate method (e.g., \inlineReviewHref{MDI0OlB1bGxSZXF1ZXN0UmV2aWV3Q29tbWVudDQ1OTk4NTYwMw==}{\textit{``Is this necessary given that super.getDependencies does the same?''}} [\inlineReviewLink{1741}]). 
We encounter this smell more often in Method (\numberToBeChecked{\waitingReview{17}} -- \numberToBeChecked{\waitingReview{twelve}} in the declaration clause and \numberToBeChecked{\waitingReview{five}} in the body block), Call (\numberToBeChecked{8}), and Attribute (\numberToBeChecked{8}). These three comprise more than half of the instances in this category.

In \numberToBeChecked{\waitingReview{10.9\%}} (\numberToBeChecked{\waitingReview{42/385}}) of the code review comments the reviewers identified \textsc{inconsistent or disrupted formatting}. This \numberToBeChecked{understandability smell} is linked to spaces, braces, parentheses, and formatting styles. The formatting elements can help the reader to identify and delimiter the elements in source code, e.g., the use of curly braces (\texttt{\{\}}) to define the scope of a block. This includes comments that identified missing vertical and horizontal space (e.g., \inlineReviewHref{MDI0OlB1bGxSZXF1ZXN0UmV2aWV3Q29tbWVudDM2OTA1MzcyMg==}{\textit{``more spaces please''}} [\inlineReviewLink{683}]); 
missing parentheses to make expression evaluation order explicit  (e.g., \inlineReviewHref{MDI0OlB1bGxSZXF1ZXN0UmV2aWV3Q29tbWVudDUwMDU2MDg3Ng==}{\textit{``Suggested change \texttt{...requiresMaintenance() \&\& null != u.getTech()} $\rightarrow$ \texttt{...requiresMaintenance() \&\& (null != u.getTech())}''}} [\inlineReviewLink{1453}]); 
and missing brackets and line breaks.
The reviewers also pointed out extraneous vertical and horizontal spaces 
and parentheses (e.g., \inlineReviewHref{MDI0OlB1bGxSZXF1ZXN0UmV2aWV3Q29tbWVudDQ0NTMwMjQ3MA==}{\textit{``A little too many parens here''}} [\inlineReviewLink{1446}]). 
Additionally, we run into comments that indicate code formatting that does not match formatting guidelines, e.g., in \texttt{if} statements,  \inlineReviewHref{MDI0OlB1bGxSZXF1ZXN0UmV2aWV3Q29tbWVudDQ2MzI1NTc2NA==}{\textit{``IF statement should be on it's own line''}} [\inlineReviewLink{2024}]).
These cases of inadequate formatting are most frequently \numberToBeChecked{\waitingReview{57.1\% (24/42)}} found in Method (\waitingReview{17} -- \numberToBeChecked{\waitingReview{seven}} in the declaration clause and \numberToBeChecked{\waitingReview{ten}} in the body) and Class (\numberToBeChecked{7}).

We find in \numberToBeChecked{\waitingReview{8.1\% (31/385)}} of our sample code review comments related to \textsc{Wrong, missing, or inadequate string expression or literal}. This \numberToBeChecked{understandability smell} pertains to concerns regarding natural language string messages such as those used as arguments in exceptions or logs, as well as scenarios involving the style and typographical correctness of string values. 
In these comments, the reviewers indicated 
incorrect words in string messages (e.g., \inlineReviewHref{MDI0OlB1bGxSZXF1ZXN0UmV2aWV3Q29tbWVudDU0MTk0MjAyOA==}{\textit{``... iterable, not iterator...''}} [\inlineReviewLink{2365}]); 
different styles of string message (e.g., \inlineReviewHref{MDI0OlB1bGxSZXF1ZXN0UmV2aWV3Q29tbWVudDUzNzc4NzkwNA==}{\textit{``not supported sounds a bit better than "unsupported"''}} [\inlineReviewLink{2287}]); 
and missing string messages (e.g., \inlineReviewHref{MDI0OlB1bGxSZXF1ZXN0UmV2aWV3Q29tbWVudDQ3MzY5NTgzNw==}{\textit{``Maybe an explanation here would be nice?''}} [\inlineReviewLink{1345}]). 
Moreover, the reviewers pointed out 
the string value in a parameter that does not match the code conventions (e.g., \inlineReviewHref{MDI0OlB1bGxSZXF1ZXN0UmV2aWV3Q29tbWVudDQzNjM0MTE2OA==}{\textit{``Suggested change \texttt{@Column(name = "END\_TEXT")} $\rightarrow$ \texttt{@Column(name = "end\_text")}''}} [\inlineReviewLink{1418}]); and a typo in a string value (\inlineReviewHref{MDI0OlB1bGxSZXF1ZXN0UmV2aWV3Q29tbWVudDUzMjQ1NjQzMA==}{\textit{``Suggested change \texttt{operationId = "aproveInboxItemById"} $\rightarrow$ \texttt{operationId = "approveInboxItemById"}}''} [\inlineReviewLink{1642}]). 
By definition, all comments where this smell manifests pertain to concerns found in existing literals during code review. \waitingReview{There are two exceptions to this pattern: when the smell is found in a throw invocation without a literal, and when the format method is used to construct a string. In the first case, we identify the Try-Catch node as the source of the smell, and in the second case, we consider the Call node.}

The instances of the \textsc{inadequate logging and monitoring} smell account for \numberToBeChecked{\waitingReview{4.7\% (18/385)}} of the code review comments. This comprises scenarios related to the logs and exceptions statements. These statements communicate to developers information about system behavior at run time, e.g., for debugging and monitoring purposes. This information can also help developers to make sense of the code functionality around these statements. 
In these comments, reviewers indicated concerns about the usage of logs and exceptions, such as 
missing log (\inlineReviewHref{MDI0OlB1bGxSZXF1ZXN0UmV2aWV3Q29tbWVudDQyMDc0NjczNw==}{\textit{``should we log this? just to make investigation easier''}} [\inlineReviewLink{1135}]); 
unnecessary log (\inlineReviewHref{MDI0OlB1bGxSZXF1ZXN0UmV2aWV3Q29tbWVudDQ2MzkwNjc1MQ==}{\textit{``In my opinion, we do not need this log. WDYT?''}} [\inlineReviewLink{2337}]); 
incoherent log levels (e.g., \inlineReviewHref{MDI0OlB1bGxSZXF1ZXN0UmV2aWV3Q29tbWVudDM3ODk3MjM2Mw==}{\textit{``...this is currently logging every second in test. Can this be changed to debug?''}} [\inlineReviewLink{574}]); 
and throwing exceptions of generic types (e.g., \inlineReviewHref{MDI0OlB1bGxSZXF1ZXN0UmV2aWV3Q29tbWVudDQ2MjMzMzMxNw==}{\textit{``This is not a very good exception, let's tell the developer what is wrong in each case and how they could fix it''}} [\inlineReviewLink{919}]). 
Most of the instances of this \numberToBeChecked{understandability smell} are related to Call (\numberToBeChecked{\waitingReview{11/18}}) statements. This smell does not cover issues with text literals used as log messages, as they are instances of the \textsc{Wrong, missing, or inadequate string expression or literal} \numberToBeChecked{smell}.

Finally, \numberToBeChecked{\waitingReview{2.3\% (9/385)}} of the code review comments are about \textsc{Missing constant usage}. Constants are often used to indicate the purpose of literals with a specific meaning in a program, e.g., using a constant named \texttt{PI} instead of its actual value. Therefore, one of their goals is to improve understandability. In these code reviews, the reviewers pointed out the direct use of literal values (e.g., \inlineReviewHref{MDI0OlB1bGxSZXF1ZXN0UmV2aWV3Q29tbWVudDQxMzgxNTk3OA==}{\textit{``It would be better to declare this Strings as constants''}} [\inlineReviewLink{2163}]), even when there are constants available (e.g., \inlineReviewHref{MDI0OlB1bGxSZXF1ZXN0UmV2aWV3Q29tbWVudDQ0NDg2NjYwMQ==}{\textit{``...You should also be able to reuse the constants declared at the beginning of \texttt{ConflictsWithScrutinizer}''}} [\inlineReviewLink{1687}]). By definition, reviewers found these \numberToBeChecked{understandability smells} in Literals.

\begin{summary}
\begin{small}
\textbf{Answer to RQ2.}
In our representative sample of \waitingReview{385} code reviews, we found \numberToBeChecked{eight} categories of understandability smells. \textsc{Missing or inadequate code documentation} (\numberToBeChecked{\waitingReview{86}}), \textsc{Bad identifier} (\numberToBeChecked{\waitingReview{78}}), and \textsc{Complex, long, or inadequate logic} (\numberToBeChecked{\waitingReview{70}}) comprise \numberToBeChecked{more than 60\%} of the understandability smells in our sample. They are related to \numberToBeChecked{16} source code element types. The three most frequent places, i.e., Method (\numberToBeChecked{\waitingReview{130}}), Call (\numberToBeChecked{\waitingReview{55}}), and Literal (\numberToBeChecked{\waitingReview{40}}), account for more than \numberToBeChecked{\waitingReview{58\%}} of the code reviews. 

\vspace{5pt}
\noindent\textbf{Implications.}
We discovered a variety of issues affecting code understandability and grouped them into \numberToBeChecked{understandability smells}. Automated solutions to improve code understandability must deal with these different kinds of issues, with varying levels of context dependence, and tackling multiple modalities: source code as well as unstructured and structured natural language.
The results suggest that tools based on LLMs have to be trained in these different languages, as some smells manifest only in Java source code (e.g., \textsc{Unnecessary code}) whereas others occur in natural language text or identifiers, e.g., \textsc{Bad identifier} and \textsc{Incomplete or inadequate code documentation}.
\end{small}
\end{summary}

\subsection{\rqthree (RQ3)} \label{results:rq3}

To answer this question, we analyzed a set of \numberToBeChecked{\waitingReview{385}} code review comments related to code understandability improvements (the same set analyzed in~\autoref{results:rq2}) and added another random sample of \numberToBeChecked{\waitingReview{385}} code review comments not related to code understandability. The code review comments were classified based on two properties: (i) whether they suggested an improvement for understandability, and (ii) whether the suggested changes were accepted or not. The results of this classification are presented in \autoref{tab:rq3:frequency}. This contingency table allows us to investigate the odds of accepting a code review comment, given that it is an understandability improvement.

We verified whether code review comments that include suggestions for improving code understandability are more likely to be accepted by developers than comments that do not include such suggestions. Our analysis revealed that code review comments that suggest understandability improvements are more likely to be accepted than comments that do not address understandability: the odds ratio is \numberToBeChecked{2.69}, with a confidence interval [\numberToBeChecked{\waitingReview{1.71, 4.22}}] for a 99\% confidence level. We also performed a Chi-Square test to evaluate the relationship between these two variables and found a significant correlation between them. In fact, suggestions for improving code understandability were found to be significantly more likely to be accepted than comments unrelated to understandability (\waitingReview{$\chi^2(1, N=440) = 32.9, p<0.01$}).

Additionally, we conducted a detailed analysis of the code reviews that suggested improvements in understandability. We observed that in \numberToBeChecked{three} instances where understandability improvements were suggested, the developers accepted that the code needed improvement, but they did not agree with the reviewer's suggestion. Instead, they implemented what they believed was the best alternative without any discussion. For instance, in one code review, the reviewer suggested \inlineReviewHref{MDI0OlB1bGxSZXF1ZXN0UmV2aWV3Q29tbWVudDQ2NjQzODUwNQ==}{\textit{``rename \texttt{module\_uuid} to \texttt{assembly\_uuid}''}} [\inlineReviewLink{1935}], but the developer renamed it to \texttt{assemblyId}. Also, in \numberToBeChecked{eight} of the analyzed code reviews developers answered that they accepted the suggestion, but we cannot find any change, even when the developer answered \inlineReviewHref{MDI0OlB1bGxSZXF1ZXN0UmV2aWV3Q29tbWVudDM3MDI5MzM5MQ==}{\textit{``changed''}} [\inlineReviewLink{1601}]. In \numberToBeChecked{\waitingReview{25}} instances of suggestions of code understandability improvements that were not accepted, the developers argued in favor of their alternative. For example, during a code review, a reviewer suggests \inlineReviewHref{MDI0OlB1bGxSZXF1ZXN0UmV2aWV3Q29tbWVudDQ2NTA5Njk5NQ==}{initializing a boolean attribute with the value \texttt{false}} [\inlineReviewLink{854}], but the developer argues that it is redundant because the default value of a boolean variable is already \texttt{false}.

Developers have given a similar level of attention to the understandability smells. For example, there were \numberToBeChecked{\waitingReview{71}} patches aiming to address \textsc{Bad identifier} (\autoref{tab:rq4:result} provides the data for the remaining smells) and \numberToBeChecked{\waitingReview{78}} instances of that smell (the rightmost column of \autoref{tab:rq2:result}). This means that \waitingReview{91\%} of the instances of this smell resulted in an accepted patch with an improvement, the highest percentage among the understandability smells. Overall, \numberToBeChecked{\waitingReview{83.9\% (323/385)}} of the code review comments triggered code understandability improvements. For \textsc{\waitingReview{Missing constant usage}}, the percentage was \waitingReview{66.7\% (6/9)}, the lowest among the smells and the most distant from the mean of \waitingReview{82.13\%}. Calculation of the odds ratio considering the groups where this latter smell was fixed vs. not fixed and where any other smell was fixed vs. not fixed suggests that the difference is not statistically significant (\waitingReview{OR=0.37, \textit{p}=0.15}).

\begin{table}[t]
	\caption{Frequency of accepted vs. not accepted code review suggestions with and without understandability improvements}
    \label{tab:rq3:frequency}
	\centering
	\scriptsize
 \vspace{-5pt}
	\begin{tabular}{@{}l @{}r r r}
		\toprule
		\multirow{2}{*}{Contingency table (770)} & & \multicolumn{2}{c}{Understandability Improvement} \\
            & & Presence & Absence \\
		\midrule
            \multirow{2}{*}{Acceptability} & Accepted & 323 & 254 \\
            & Not accepted & 62 & 131 \\
		\bottomrule
	\end{tabular}
 \vspace{-13pt}
\end{table}

\begin{summary}
\begin{small}
\textbf{Answer to RQ3.}
Code review comments that suggest understandability improvements are significantly more likely to be accepted than comments that do not address understandability improvements. The majority (\numberToBeChecked{\waitingReview{83.9\%}}) of the suggestions for code understandability improvements in code review comments were accepted. \waitingReview{Less than \numberToBeChecked{1\%}} acknowledged the need for improvement but applied another solution. 
However, in  \numberToBeChecked{\waitingReview{16.1\%}} of the cases, the pull request authors did not accept the suggestion. In addition, the smells received similar levels of attention, in terms of the percentage of instances for which an understandability improvement was accepted. 

\vspace{5pt}
\noindent\textbf{Implications.}
Developers typically accept suggestions for improving code understandability during the code review process. These suggestions and the changes related to them can be valuable and trustworthy data for devising automated approaches to improve code understandability. \waitingReview{For example, they can support few-shot learning when leveraging LLMs to make code more readable.}
\end{small}
\end{summary}

\subsection{\rqfour (RQ4)} \label{results:rq4}

We examined \numberToBeChecked{\waitingReview{323}} code review comments from the results of our analysis in \autoref{results:rq3} to assess improvements in code understandability. We sorted and categorized the improvements made to address the understandability smells in these reviews, and the results are presented in \autoref{tab:rq4:result}.  The patches are organized based on the \numberToBeChecked{understandability smells} presented in \autoref{results:rq2}, e.g., \textsc{Unnecessary code} and \textsc{Bad identifier}. The similar patches in each category were organized in groups, e.g., $\hookrightarrow$ \underline{Unused code}. In \autoref{tab:rq4:result}, an arrow ($\hookrightarrow$) represents a group and its name is underlined. We also combined the patches based on the action taken by the developer and we created a second level of indentation on the table to avoid repetition of text, e.g., \underline{explain the functionality of a method}. Patches appearing three or more times are marked with the clubsuit symbol ($\clubsuit$) to indicate their higher frequency.
In the remainder of this section, we discuss the understandability improvements that were applied for each understandability smell.

\vspace{5pt}
\noindent\textbf{Incomplete or inadequate code documentation}. We found \numberToBeChecked{\waitingReview{71}} patches that added, removed, enhanced, or fixed Javadoc and code comments in the source code. The majority of them (\numberToBeChecked{\waitingReview{15}}) are patches that improved the grammar or fixed a typo in existing Javadoc (e.g., \inlineReviewHref{MDI0OlB1bGxSZXF1ZXN0UmV2aWV3Q29tbWVudDM5OTQxOTAxNw==}{\textit{``memory which available'' $\rightarrow$ ``memory which is available'}} [\inlineReviewLink{1115}]). Also  \numberToBeChecked{\waitingReview{nine}} patches added  Javadoc documentation to a method or class where it was absent (e.g., \inlineReviewHref{MDI0OlB1bGxSZXF1ZXN0UmV2aWV3Q29tbWVudDQ0OTMyNzEwNQ==}{methods} [\inlineReviewLink{756}] and \inlineReviewHref{MDI0OlB1bGxSZXF1ZXN0UmV2aWV3Q29tbWVudDQ3NTg3NzU1Ng==}{classes} [\inlineReviewLink{2244}]). In other patches related to Javadoc, the developers enhanced the description of code elements (e.g., \inlineReviewHref{MDI0OlB1bGxSZXF1ZXN0UmV2aWV3Q29tbWVudDQwMjA3NjU4NA==}{\textit{``Time spent for waiting a task to be completed.'' $\rightarrow$ ``This time spent while waiting for the task to be completed is being recorded in this counter.''}} [\inlineReviewLink{151}]) and added tags and hyperlinks (e.g., \inlineReviewHref{MDI0OlB1bGxSZXF1ZXN0UmV2aWV3Q29tbWVudDQ3NzAwOTA3OA==}{\textit{``{@link JobModel}''}} [\inlineReviewLink{294}]). In terms of changes applied to code comments, there are patches that \inlineReviewHref{MDI0OlB1bGxSZXF1ZXN0UmV2aWV3Q29tbWVudDQ1NDcwMDUwNg==}{added code comments to explain the choice of an attribute type} [\inlineReviewLink{2180}], \inlineReviewHref{MDI0OlB1bGxSZXF1ZXN0UmV2aWV3Q29tbWVudDM3MDkyMzYzMA==}{removed TODO code comments for completed tasks} [\inlineReviewLink{2387}], and \inlineReviewHref{MDI0OlB1bGxSZXF1ZXN0UmV2aWV3Q29tbWVudDUzNTc1NTQ4Mg==}{transformed a code comment into Javadoc} [\inlineReviewLink{1625}]. Finally, \numberToBeChecked{\waitingReview{four}} patches  \inlineReviewHref{MDI0OlB1bGxSZXF1ZXN0UmV2aWV3Q29tbWVudDQ0Njg1MjAwMA==}{added or removed annotation related to documentation, e.g., \texttt{Deprecated}} [\inlineReviewLink{2263}].

\vspace{5pt}
\noindent\textbf{Bad identifier}. \numberToBeChecked{\waitingReview{71}} patches changed an identifier to better communicate what it means in the source code. These patches involved changes regarding both the content and style of the identifiers. To address concerns related to the content of the identifiers, the majority (\numberToBeChecked{\waitingReview{50}}) of the patches focused on clarifying their meaning (e.g., \inlineReviewHref{MDI0OlB1bGxSZXF1ZXN0UmV2aWV3Q29tbWVudDUwNzE3NDQxMQ==}{\texttt{String getDates} $\rightarrow$ \texttt{String getDateRange}} [\inlineReviewLink{1042}]) and expressing the identifier's type (e.g., \inlineReviewHref{MDI0OlB1bGxSZXF1ZXN0UmV2aWV3Q29tbWVudDUxNzMwNTQxNw==}{\texttt{RequestExecutor requestHandler} $\rightarrow$ \texttt{RequestExecutor requestExecutor}} [\inlineReviewLink{340}]). Additionally, \numberToBeChecked{\waitingReview{eleven}} patches fixed typos (e.g., \inlineReviewHref{MDI0OlB1bGxSZXF1ZXN0UmV2aWV3Q29tbWVudDUyMDgzNjQ1OA==}{\texttt{verifyEntityTypeMigrationInValidEntities} $\rightarrow$ \texttt{verifyEntityTypeMigrationInvalidEntities}} [\inlineReviewLink{1107}]) and \numberToBeChecked{four} adjusted the identifiers to adhere to the conventions adopted by the project (e.g., \inlineReviewHref{MDI0OlB1bGxSZXF1ZXN0UmV2aWV3Q29tbWVudDQ2MDc0NDA1Ng==}{\texttt{boolean isActive} $\rightarrow$ \texttt{boolean active}} [\inlineReviewLink{453}]). Furthermore, we observed \numberToBeChecked{\waitingReview{six}} patches that aimed to improve the style of the identifiers, such as converting them to \inlineReviewHref{MDI0OlB1bGxSZXF1ZXN0UmV2aWV3Q29tbWVudDQ0NTgyNzc0OQ==}{lowercase 
[\inlineReviewLink{579}]} or \inlineReviewHref{MDI0OlB1bGxSZXF1ZXN0UmV2aWV3Q29tbWVudDQ5MDI0MzE0Mg==}{camelcase 
[\inlineReviewLink{1753}]}.

\vspace{5pt}
\noindent\textbf{Complex, long, or inadequate logic}. \numberToBeChecked{\waitingReview{54}} patches extracted, moved, or replaced complex, lengthy, or inconsistent code in an attempt to simplify it. Among these, five patches extracted parts of a method to reduce its length (e.g., \inlineReviewHref{MDI0OlB1bGxSZXF1ZXN0UmV2aWV3Q29tbWVudDQ5ODgyNjU5Mg==}{extracting a method}).
Also, developers applied \numberToBeChecked{\waitingReview{16}} patches that replaced a usage of an external API or adopted an external API to simplify complex code, such as \inlineReviewHref{MDI0OlB1bGxSZXF1ZXN0UmV2aWV3Q29tbWVudDQ2MDY4NjQyNQ==}{replacing a chain of method invocations by a unique API method invocation} [\inlineReviewLink{1700}]. In 
\numberToBeChecked{\waitingReview{14}} other scenarios, developers replaced expressions in source code by an alternative that better communicates the functionality of the code, e.g., \inlineReviewHref{MDI0OlB1bGxSZXF1ZXN0UmV2aWV3Q29tbWVudDM3MzI4NTI4MA==}{replacing boolean expression by a call to method   \texttt{isEmpty()}} [\inlineReviewLink{626}]. Additionally, developers applied patches that replace the use of a construct in \numberToBeChecked{\waitingReview{eight}} cases, such as \inlineReviewHref{MDI0OlB1bGxSZXF1ZXN0UmV2aWV3Q29tbWVudDQyNDYwODY0Nw==}{replacing an anonymous inner class by a lambda} [\inlineReviewLink{1971}] or \inlineReviewHref{MDI0OlB1bGxSZXF1ZXN0UmV2aWV3Q29tbWVudDUyNzQ4NDAzMQ==}{replacing a lambda with conditional by an \texttt{if} statement} [\inlineReviewLink{1892}]. Finally, the developers applied \numberToBeChecked{\waitingReview{five}} patches to organize the use of APIs in a source code file, such as expanding imports instead of using wildcards [\inlineReviewLink{2203}], and adding an \texttt{import} statement to avoid the use of the fully-qualified name of a class (\inlineReviewHref{MDI0OlB1bGxSZXF1ZXN0UmV2aWV3Q29tbWVudDQzMTYyNDA2OA==}{\texttt{ TileSourceManager.TileSourceTemplate template} $\rightarrow$ \texttt{import ...TileSourceManager; private TileSourceTemplate template}} [\inlineReviewLink{1779}]).

\begin{table}[t!]
    \caption{Understandability smell categories and the improvement patches found in the \waitingReview{323} accepted code reviews. The patches with three or more instances are marked with the symbol $\clubsuit$. In the PDF version, each patch links to an example.}
    \label{tab:rq4:result}
    \scriptsize
\hspace{-20pt}
\vspace{-5pt}
\begin{tabular}{l}
        \toprule
        \textbf{Category} $\hookrightarrow$ \underline{Group} or type of improvement patches (\# found) \\
        \midrule
        
        \rowcolor{gray!15}\textbf{Incomplete or inadequate code documentation \waitingReview{(71)}} \\
            $\hookrightarrow$ \underline{Changes related to Javadoc}: \waitingReview{(50)} \\
                \tabindent -- \inlineReviewHref{MDI0OlB1bGxSZXF1ZXN0UmV2aWV3Q29tbWVudDUyNzEyOTQxOA==}{Add new Javadoc block to describe the functionality of a code element} \waitingReview{(9)} $\clubsuit$ \\
                \tabindent -- \inlineReviewHref{MDI0OlB1bGxSZXF1ZXN0UmV2aWV3Q29tbWVudDQ3NzAwOTA3OA==}{Add new tag and hyperlink to existing Javadoc} \waitingReview{(3) $\clubsuit$} \\
                \tabindent -- Add text to existing Javadoc to \\
                    \tabindent\tabindent - \inlineReviewHref{MDI0OlB1bGxSZXF1ZXN0UmV2aWV3Q29tbWVudDM5ODY1MDc3Nw==}{explain the functionality of a method} (5) $\clubsuit$\\
                    \tabindent\tabindent - \inlineReviewHref{MDI0OlB1bGxSZXF1ZXN0UmV2aWV3Q29tbWVudDQ4MTc1MDgwMg==}{explain internal implementation details of a method} (2) \\
                    \tabindent\tabindent - \inlineReviewHref{MDI0OlB1bGxSZXF1ZXN0UmV2aWV3Q29tbWVudDU0OTgxNDkyNA==}{improve explanation about code element} (2) \\
                    \tabindent\tabindent - \inlineReviewHref{MDI0OlB1bGxSZXF1ZXN0UmV2aWV3Q29tbWVudDUxMjcxMTIwOQ==}{explain reason for deprecation} (1) \\
                \tabindent -- Change text in existing Javadoc to\\
                    \tabindent\tabindent - \inlineReviewHref{MDI0OlB1bGxSZXF1ZXN0UmV2aWV3Q29tbWVudDQ1MzMzMzI0MQ==}{improve grammar or fix a typo} \waitingReview{(15)} $\clubsuit$ \\
                    \tabindent\tabindent -                \inlineReviewHref{MDI0OlB1bGxSZXF1ZXN0UmV2aWV3Q29tbWVudDQwMjA3NjU4NA==}{improve explanation about functionality of a method} (6) $\clubsuit$ \\
                    \tabindent\tabindent - \inlineReviewHref{MDI0OlB1bGxSZXF1ZXN0UmV2aWV3Q29tbWVudDQ2Mzc3Mzg4MQ==}{update copyright} \waitingReview{(3) $\clubsuit$} \\
                    \tabindent\tabindent - \inlineReviewHref{MDI0OlB1bGxSZXF1ZXN0UmV2aWV3Q29tbWVudDQ0Mzc0MDIxNQ==}{better explain the purpose of an annotation} (1) \\
                    \tabindent\tabindent - \inlineReviewHref{MDI0OlB1bGxSZXF1ZXN0UmV2aWV3Q29tbWVudDQ5OTQ3NDA5MQ==}{fix the name of a code element to which it refers} (1) \\
                    \tabindent\tabindent - \inlineReviewHref{MDI0OlB1bGxSZXF1ZXN0UmV2aWV3Q29tbWVudDQ1ODMyMDY1OA==}{make it more concise} (1) \\
                \tabindent -- \inlineReviewHref{MDI0OlB1bGxSZXF1ZXN0UmV2aWV3Q29tbWVudDM4NzczMTM4Mg==}{Move a tag of documentation to the conventional place} (1) \\
            
            $\hookrightarrow$ \underline{Changes related to code comments}: \waitingReview{(17)} \\
                \tabindent -- Add code comment to\\
                \tabindent\tabindent -                \inlineReviewHref{MDI0OlB1bGxSZXF1ZXN0UmV2aWV3Q29tbWVudDM5MDYzNTEyMg==}{explain an expression} \waitingReview{(3) $\clubsuit$} \\
                \tabindent\tabindent -  \inlineReviewHref{MDI0OlB1bGxSZXF1ZXN0UmV2aWV3Q29tbWVudDQzNTY5MTcwMA==}{explain each part of a test} (2) \\
                \tabindent\tabindent - \inlineReviewHref{MDI0OlB1bGxSZXF1ZXN0UmV2aWV3Q29tbWVudDQ1NDcwMDUwNg==}{explain the type of an attribute} (1) \\
                \tabindent -- \inlineReviewHref{MDI0OlB1bGxSZXF1ZXN0UmV2aWV3Q29tbWVudDM3MDkyMzYzMA==}{Remove TODO code comment for tasks that have already been performed} (2) \\
                \tabindent -- \inlineReviewHref{MDI0OlB1bGxSZXF1ZXN0UmV2aWV3Q29tbWVudDM2NzkxNzU5Nw==}{Remove a useless code comment} (2) \\
                \tabindent -- \inlineReviewHref{MDI0OlB1bGxSZXF1ZXN0UmV2aWV3Q29tbWVudDQwNjQ1NTAzMA==}{Change code comment to better explain the associated code element} \waitingReview{(5) $\clubsuit$} \\
                \tabindent -- \inlineReviewHref{MDI0OlB1bGxSZXF1ZXN0UmV2aWV3Q29tbWVudDUyNjU3Mzk0MA==}{Change text in code comment to fix typo} (1) \\
                \tabindent -- \inlineReviewHref{MDI0OlB1bGxSZXF1ZXN0UmV2aWV3Q29tbWVudDUzNTc1NTQ4Mg==}{Transform a code comment into Javadoc documentation} (1) \\

            $\hookrightarrow$ \waitingReview{\inlineReviewHref{MDI0OlB1bGxSZXF1ZXN0UmV2aWV3Q29tbWVudDQ0Njg1MjAwMA==}{Add or remove annotation related to documentation} (4)} $\clubsuit$\\
            
        \midrule
        \rowcolor{gray!15}\textbf{Bad identifier \waitingReview{(71)}} \\
            $\hookrightarrow$ \underline{Changes related to content}: \waitingReview{(65)} \\
                \tabindent -- \inlineReviewHref{MDI0OlB1bGxSZXF1ZXN0UmV2aWV3Q29tbWVudDQwODA3MjA2OA==}{Modify an identifier to express the meaning or type of an element} \waitingReview{(50)} $\clubsuit$\\
                \tabindent -- \inlineReviewHref{MDI0OlB1bGxSZXF1ZXN0UmV2aWV3Q29tbWVudDUyMTkzNjAzOQ==}{Modify an identifier to fix a typo} \waitingReview{(11)} $\clubsuit$\\
                \tabindent -- \inlineReviewHref{MDI0OlB1bGxSZXF1ZXN0UmV2aWV3Q29tbWVudDQ2MDc0NDA1Ng==}{Modify an identifier to be consistent with a convention} \waitingReview{(4)} $\clubsuit$\\

            $\hookrightarrow$ \underline{Changes related to style}: \waitingReview{(6)}\\
                \tabindent -- \inlineReviewHref{MDI0OlB1bGxSZXF1ZXN0UmV2aWV3Q29tbWVudDU0OTMwNTAwMA==}{Change the style of an identifier to be camelCase, capitalized, or lowercase} \waitingReview{(6)} $\clubsuit$\\

        \midrule

        \rowcolor{gray!15}\textbf{Complex, long, or inadequate logic \waitingReview{(54)}} \\
            $\hookrightarrow$ \inlineReviewHref{MDI0OlB1bGxSZXF1ZXN0UmV2aWV3Q29tbWVudDQ5NDQ0MTY1Mw==}{Extract method} \waitingReview{(6)} $\clubsuit$\\
            $\hookrightarrow$ \inlineReviewHref{MDI0OlB1bGxSZXF1ZXN0UmV2aWV3Q29tbWVudDQzMzE1Mjc2OQ==}{Extract variable} (1) \\
            $\hookrightarrow$ \inlineReviewHref{MDI0OlB1bGxSZXF1ZXN0UmV2aWV3Q29tbWVudDQ0MTQ3NzA2NQ==}{Rewrite the code to avoid semantic duplicate method call} \waitingReview{(3) $\clubsuit$} \\
            $\hookrightarrow$ \inlineReviewHref{MDI0OlB1bGxSZXF1ZXN0UmV2aWV3Q29tbWVudDUxNzE1MTM3NQ==}{Move the code element to consistent place} (1) \\
            
            $\hookrightarrow$ \underline{Changes related to usage of APIs}: \waitingReview{(16)} \\
                \tabindent -- \inlineReviewHref{MDI0OlB1bGxSZXF1ZXN0UmV2aWV3Q29tbWVudDM4NTI3MTU5Mg==}{Replace method calling chain by existing API} (6) $\clubsuit$\\
                \tabindent -- \inlineReviewHref{MDI0OlB1bGxSZXF1ZXN0UmV2aWV3Q29tbWVudDUyOTk2OTI5NQ==}{Use a different type to be consistent with other parts of the code} \waitingReview{(5)} $\clubsuit$\\
                \tabindent -- \inlineReviewHref{MDI0OlB1bGxSZXF1ZXN0UmV2aWV3Q29tbWVudDM5ODM1ODU1Mw==}{Replace usage of an external API by an internal one} (2) \\
                \tabindent -- \inlineReviewHref{MDI0OlB1bGxSZXF1ZXN0UmV2aWV3Q29tbWVudDQ1MTAwOTA3OA==}{Replace a set of statements by direct method in existing API} (2) \\
                \tabindent -- \inlineReviewHref{MDI0OlB1bGxSZXF1ZXN0UmV2aWV3Q29tbWVudDM5NTk5MzQyNA==}{Replace Lists.newArrayList by Arrays.asList} (1) \\
            
            $\hookrightarrow$ \underline{Changes related to expressions}: \waitingReview{(14)} \\
                
                \tabindent -- \inlineReviewHref{MDI0OlB1bGxSZXF1ZXN0UmV2aWV3Q29tbWVudDM3NTAyMjI2Nw==}{Replace an expression by an alternative expression} \waitingReview{(5) $\clubsuit$} \\
                \tabindent -- \inlineReviewHref{MDI0OlB1bGxSZXF1ZXN0UmV2aWV3Q29tbWVudDQ3Njc2NDg4NA==}{Replace call to static method by instance method call} (3) $\clubsuit$\\
                \tabindent -- \inlineReviewHref{MDI0OlB1bGxSZXF1ZXN0UmV2aWV3Q29tbWVudDM3MzI4NTI4MA==}{Replace boolean expression by an alternative direct method} (2) \\
                \tabindent -- \inlineReviewHref{MDI0OlB1bGxSZXF1ZXN0UmV2aWV3Q29tbWVudDQwMjMwNjQ0MQ==}{Replace the use of a parameter by an expression with another parameter} (1) \\
                \tabindent -- \inlineReviewHref{MDI0OlB1bGxSZXF1ZXN0UmV2aWV3Q29tbWVudDQ2MzY5NTYwOQ==}{Replace indirect expression in parameter by direct value} (1) \\
                \tabindent -- \inlineReviewHref{MDI0OlB1bGxSZXF1ZXN0UmV2aWV3Q29tbWVudDU2MjE4MzMzNg==}{Replace an expression that creates object by existing attribute} (1) \\
                \tabindent -- \inlineReviewHref{MDI0OlB1bGxSZXF1ZXN0UmV2aWV3Q29tbWVudDUyMjQwNzQ3Mw==}{Invert the boolean expression to simplify it} (1) \\

            $\hookrightarrow$ \underline{Changes related to constructs}: \waitingReview{(8)} \\

                \tabindent -- \inlineReviewHref{MDI0OlB1bGxSZXF1ZXN0UmV2aWV3Q29tbWVudDQwNjM1OTE1MQ==}{Replace an imperative logic by functional} \waitingReview{(3) $\clubsuit$} \\
                \tabindent -- \inlineReviewHref{MDI0OlB1bGxSZXF1ZXN0UmV2aWV3Q29tbWVudDU1NjU4NDEyNw==}{Replace anonymous inner class or method calling chain by lambda} (3) $\clubsuit$ \\
                \tabindent -- \inlineReviewHref{MDI0OlB1bGxSZXF1ZXN0UmV2aWV3Q29tbWVudDQzNzY4NTg0MQ==}{Replace the while and switch structure by an alternative switch with default} (1) \\
                \tabindent -- \inlineReviewHref{MDI0OlB1bGxSZXF1ZXN0UmV2aWV3Q29tbWVudDQ5MTEzODQ4NA==}{Replace short-circuit in stream by an alternative reducing operation} (1) \\
              
            $\hookrightarrow$ \underline{Changes related to imports}: \waitingReview{(5)} \\
            
                \tabindent -- \inlineReviewHref{MDI0OlB1bGxSZXF1ZXN0UmV2aWV3Q29tbWVudDUzOTI0MTU2Mg==}{Expand the imports} \waitingReview{(3) $\clubsuit$} \\
                \tabindent -- \inlineReviewHref{MDI0OlB1bGxSZXF1ZXN0UmV2aWV3Q29tbWVudDQzMTYyNDA2OA==}{Remove the library name reference in fully qualifier name} \waitingReview{(2)} \\
        
        \midrule
        
        \multicolumn{1}{r}{\textit{Continued on the next column}} \\

        \bottomrule
        
    \end{tabular}
    \vspace{-8pt}
\end{table}

\begin{table}[h!]
    \scriptsize

\begin{tabular}{l}
        \textit{Continued from the previous column} \\
        \toprule
        \textbf{Category} $\hookrightarrow$ \underline{Group} or type of improvement patches (\# found) \\
        \midrule

        \rowcolor{gray!15}\textbf{Unnecessary code \waitingReview{(43)}} \\
            $\hookrightarrow$ \underline{Unused code}: \waitingReview{(24)} \\
            \tabindent -- \inlineReviewHref{MDI0OlB1bGxSZXF1ZXN0UmV2aWV3Q29tbWVudDU0NzIzNDcwMA==}{Remove non-referred constant, import, class, method, or variable} \waitingReview{(19)} $\clubsuit$\\

            \tabindent -- \inlineReviewHref{MDI0OlB1bGxSZXF1ZXN0UmV2aWV3Q29tbWVudDU0Njc4NDIyOQ==}{Remove commented out code statements} (5) $\clubsuit$\\

            $\hookrightarrow$ \underline{Redundant code}: \waitingReview{(19)} \\

            \tabindent -- \inlineReviewHref{MDI0OlB1bGxSZXF1ZXN0UmV2aWV3Q29tbWVudDQ1OTk4NTYwMw==}{Remove duplicated code or processing} \waitingReview{(13)} $\clubsuit$\\
            
            \tabindent -- \inlineReviewHref{MDI0OlB1bGxSZXF1ZXN0UmV2aWV3Q29tbWVudDUzMzMyNjg4NQ==}{Remove null constants} (1) \\

            \tabindent -- \inlineReviewHref{MDI0OlB1bGxSZXF1ZXN0UmV2aWV3Q29tbWVudDUyODU2NTU2NA==}{Remove type parameter for inferrable type} (1) \\

            \tabindent -- \inlineReviewHref{MDI0OlB1bGxSZXF1ZXN0UmV2aWV3Q29tbWVudDU1NjcwODAyNw==}{Remove \texttt{this}} (1) \\
            
            \tabindent -- \inlineReviewHref{MDI0OlB1bGxSZXF1ZXN0UmV2aWV3Q29tbWVudDQ4NDUzMTYzOA==}{Remove imported type from fully qualified names} (1) \\
            
            \tabindent -- \inlineReviewHref{MDI0OlB1bGxSZXF1ZXN0UmV2aWV3Q29tbWVudDQyNDcyOTc0Mw==}{Inline temporary} (2) \\
        \midrule
        \rowcolor{gray!15}\textbf{Inconsistent or disrupted formatting \waitingReview{(37)}}\\
            $\hookrightarrow$ \underline{Space usage}: \waitingReview{(26)} \\
                \tabindent --  \inlineReviewHref{MDI0OlB1bGxSZXF1ZXN0UmV2aWV3Q29tbWVudDU3NjgzMTU0Nw==}{Add or remove horizontal spacing} \waitingReview{(13)} $\clubsuit$\\
                \tabindent -- \inlineReviewHref{MDI0OlB1bGxSZXF1ZXN0UmV2aWV3Q29tbWVudDUwNjEyMTMxMg==}{Add or remove vertical spacing} \waitingReview{(13)} $\clubsuit$\\

            $\hookrightarrow$ \underline{Formatting element visibility}: (3) \\
                \tabindent -- \inlineReviewHref{MDI0OlB1bGxSZXF1ZXN0UmV2aWV3Q29tbWVudDM5Mjg1NjgyNw==}{Add braces in a single statement block} (2) \\
                \tabindent -- \inlineReviewHref{MDI0OlB1bGxSZXF1ZXN0UmV2aWV3Q29tbWVudDUwMDU2MDg3Ng==}{Add parentheses to compound logical expressions} (1) \\

            $\hookrightarrow$ \underline{Formatting style}: (8) \\
                \tabindent -- \inlineReviewHref{MDI0OlB1bGxSZXF1ZXN0UmV2aWV3Q29tbWVudDM4Njg2MDE4Mw==}{Add line break to a long statement line} (1) \\
                \tabindent -- \notsotiny\inlineReviewHref{MDI0OlB1bGxSZXF1ZXN0UmV2aWV3Q29tbWVudDQ0NTMwMjQ3MA==}{Remove parentheses in logical expression operands and move them to separated lines} (1) \\
                \tabindent -- \inlineReviewHref{MDI0OlB1bGxSZXF1ZXN0UmV2aWV3Q29tbWVudDM3OTkyNjA5Nw==}{Change order of boolean expression operands to make \texttt{null} the second one} (1) \\
                \tabindent -- \inlineReviewHref{MDI0OlB1bGxSZXF1ZXN0UmV2aWV3Q29tbWVudDUwODc4OTY5Nw==}{Move code elements to their own line} (3) $\clubsuit$\\
                \tabindent -- \notsotiny\inlineReviewHref{MDI0OlB1bGxSZXF1ZXN0UmV2aWV3Q29tbWVudDQxMTg1NjIwMQ==}{Move \texttt{catch} declaration to the same line of its associated \texttt{try}'s closing block brace} (1) \\
                \tabindent -- \inlineReviewHref{MDI0OlB1bGxSZXF1ZXN0UmV2aWV3Q29tbWVudDUzNDE0NDYzMg==}{Move methods to be in call ordering} (1) \\
            
        \midrule
        \rowcolor{gray!15}\textbf{Wrong, missing, or inadequate string expression or literal \waitingReview{(26)}} \\
            $\hookrightarrow$ \inlineReviewHref{MDI0OlB1bGxSZXF1ZXN0UmV2aWV3Q29tbWVudDQ3MzY5NTgzNw==}{Add a message to an exception} \waitingReview{(2)} \\
            \waitingReview{$\hookrightarrow$ \inlineReviewHref{MDI0OlB1bGxSZXF1ZXN0UmV2aWV3Q29tbWVudDUzMzM1MDM5MQ==}{Replace use of the \texttt{format} method by string concatenation} (1)} \\
            $\hookrightarrow$ \underline{Changes related to a natural language string literal}: (18) \\
                \tabindent -- \inlineReviewHref{MDI0OlB1bGxSZXF1ZXN0UmV2aWV3Q29tbWVudDQyODk2NzYwNw==}{Fix an incorrect string literal} (6) $\clubsuit$\\
                \tabindent -- \inlineReviewHref{MDI0OlB1bGxSZXF1ZXN0UmV2aWV3Q29tbWVudDQwNjg2ODg4Ng==}{Replace a string literal by a different meaning or a synonymous} (6) $\clubsuit$\\
                \tabindent -- \inlineReviewHref{MDI0OlB1bGxSZXF1ZXN0UmV2aWV3Q29tbWVudDQ0NDg1NzQ5MA==}{Extend a string literal} (3) $\clubsuit$\\
                \tabindent -- \inlineReviewHref{MDI0OlB1bGxSZXF1ZXN0UmV2aWV3Q29tbWVudDM5MzYwMDQ0MQ==}{Change literal to adhere to project standards} (3) $\clubsuit$\\
            $\hookrightarrow$ \underline{Changes related to a string magic value}: (5) \\
                \tabindent -- \inlineReviewHref{MDI0OlB1bGxSZXF1ZXN0UmV2aWV3Q29tbWVudDUzMjQ1NjQzMA==}{Fix an incorrect value} (2) \\
                \tabindent -- \inlineReviewHref{MDI0OlB1bGxSZXF1ZXN0UmV2aWV3Q29tbWVudDQzNjM0MTE2OA==}{Change magic value to adhere to project standards} (3) $\clubsuit$\\

        \midrule

        \rowcolor{gray!15}\textbf{Inadequate logging and monitoring \waitingReview{(15)}} \\
            $\hookrightarrow$ \inlineReviewHref{MDI0OlB1bGxSZXF1ZXN0UmV2aWV3Q29tbWVudDQxNTI3MDY3NA==}{Add logging code} \waitingReview{(3) $\clubsuit$} \\
            $\hookrightarrow$ \inlineReviewHref{MDI0OlB1bGxSZXF1ZXN0UmV2aWV3Q29tbWVudDQ0MTE0NDIyOA==}{Remove log} \waitingReview{(6)} $\clubsuit$\\
            $\hookrightarrow$ \inlineReviewHref{MDI0OlB1bGxSZXF1ZXN0UmV2aWV3Q29tbWVudDU5ODU0MTI5NQ==}{Change log settings} \waitingReview{(5) $\clubsuit$} \\
            $\hookrightarrow$ \inlineReviewHref{MDI0OlB1bGxSZXF1ZXN0UmV2aWV3Q29tbWVudDQ2MjMzMzMxNw==}{Change generic exception to a specific} (1) \\

        \midrule
        \rowcolor{gray!15}\textbf{Missing constant usage \waitingReview{(6)}} \\
            $\hookrightarrow$ \inlineReviewHref{MDI0OlB1bGxSZXF1ZXN0UmV2aWV3Q29tbWVudDQ2MjUzNDE2Mw==}{Replace hardcoded value by a new constant} \waitingReview{(4)} $\clubsuit$\\
            $\hookrightarrow$ \inlineReviewHref{MDI0OlB1bGxSZXF1ZXN0UmV2aWV3Q29tbWVudDUxMjkyNzc5Mw==}{Use a constant from a library} (2) \\

        \midrule
    \end{tabular}
    \vspace{-8pt}
\end{table}

\vspace{5pt}
\noindent\textbf{Unnecessary Code}. A total of \numberToBeChecked{\waitingReview{43}} patches eliminated code that has no impact on the execution. Patches that removed unused code include \numberToBeChecked{\waitingReview{24}} instances of code elements that are not being referred to (e.g., \inlineReviewHref{MDI0OlB1bGxSZXF1ZXN0UmV2aWV3Q29tbWVudDU0NzIzNDcwMA==}{\textit{``unused constant.''}}) and \numberToBeChecked{five} instances of \inlineReviewHref{MDI0OlB1bGxSZXF1ZXN0UmV2aWV3Q29tbWVudDQ2NTQxMDgzMw==}{commented out code} [\inlineReviewLink{444}]. Also, the developers applied patches that removed duplicated code (e.g., \inlineReviewHref{MDI0OlB1bGxSZXF1ZXN0UmV2aWV3Q29tbWVudDM4ODUzMjE0MQ==}{equivalent methods} [\inlineReviewLink{325}]) or processing (e.g., \inlineReviewHref{MDI0OlB1bGxSZXF1ZXN0UmV2aWV3Q29tbWVudDQ2ODA4ODUyMQ==}{iteration happening twice on the same list} [\inlineReviewLink{536}]) \numberToBeChecked{\waitingReview{13}} times, and different constants with value \texttt{null} 
[\inlineReviewLink{949}]. 
There are also \numberToBeChecked{two} patches that {\inlineReviewHref{MDI0OlB1bGxSZXF1ZXN0UmV2aWV3Q29tbWVudDM3MDU2NDEyNQ==}{inline temporary variables} [\inlineReviewLink{1561}].

\vspace{5pt}
\noindent\textbf{Inconsistent or disrupted formatting}. \numberToBeChecked{\waitingReview{37}} patches organized and standardized code formatting. These patches manage the use of formatting characters such as space and braces, as well as formatting styles. In \numberToBeChecked{\waitingReview{26}} of the patches, the developers added space between different code elements (e.g., \inlineReviewHref{MDI0OlB1bGxSZXF1ZXN0UmV2aWV3Q29tbWVudDQzNDEyNTI2MQ==}{\textit{horizontal space}} [\inlineReviewLink{493}] and \inlineReviewHref{MDI0OlB1bGxSZXF1ZXN0UmV2aWV3Q29tbWVudDQ1ODMxMjA5OQ==}{\textit{vertical space}} [\inlineReviewLink{2050}]); and removed space between related code statements (e.g., \inlineReviewHref{MDI0OlB1bGxSZXF1ZXN0UmV2aWV3Q29tbWVudDQ1NTI1NTI1NQ==}{\textit{for declaration and the first statement of the block}} [\inlineReviewLink{1660}]) or extra space between code elements in the same expression (e.g., \inlineReviewHref{MDI0OlB1bGxSZXF1ZXN0UmV2aWV3Q29tbWVudDQ5MDM2NzQyOA==}{\textit{between the point and the invoked method in an expression}} [\inlineReviewLink{1754}]). Additionally, developers added other structuring characters (e.g., braces and parentheses) to highlight the region of a code element (e.g., \inlineReviewHref{MDI0OlB1bGxSZXF1ZXN0UmV2aWV3Q29tbWVudDM5Mjg1NjgyNw==}{\textit{block}} [\inlineReviewLink{1078}] or \inlineReviewHref{MDI0OlB1bGxSZXF1ZXN0UmV2aWV3Q29tbWVudDUwMDU2MDg3Ng==}{\textit{operand}} [\inlineReviewLink{1453}]). Finally, in \numberToBeChecked{eight} cases developers changed the order or moved code elements aiming to apply a formatting style, such as keeping the code elements in their own lines (\inlineReviewHref{MDI0OlB1bGxSZXF1ZXN0UmV2aWV3Q29tbWVudDQ2MzI1NTc2NA==}{e.g., \textit{body of \texttt{if} statement in its own line}} [\inlineReviewLink{2024}]) and arranging method declarations 
[\inlineReviewLink{1860}]. 

\vspace{5pt}
\noindent\textbf{Wrong, missing, or inadequate string expression or literal}. \numberToBeChecked{\waitingReview{26}} patches addressed concerns related to incorrect, missing, or inadequate string expressions or literals. These patches involve adding, fixing, extending, or changing strings to better suit the code context and improve code understandability. In \numberToBeChecked{six} of these patches, the developers replaced a natural language string literal by an alternative with a different meaning (e.g., \inlineReviewHref{MDI0OlB1bGxSZXF1ZXN0UmV2aWV3Q29tbWVudDQ0NDY2NTQ4Nw==}{\textit{assertEquals(``not\_equals''...} $\rightarrow$ \textit{assertEquals(``Expected and actual values should be the same!''...}} [\inlineReviewLink{786}]); or a synonym (\inlineReviewHref{MDI0OlB1bGxSZXF1ZXN0UmV2aWV3Q29tbWVudDUzNzc4NzkwNA==}{\textit{...are unsupported} $\rightarrow$ \textit{...not supported}} [\inlineReviewLink{2287}]). Furthermore, in \numberToBeChecked{six} other patches developers fixed incorrect words in string expressions (\inlineReviewHref{MDI0OlB1bGxSZXF1ZXN0UmV2aWV3Q29tbWVudDQ5NjcxODY0NA==}{\textit{Can not set min...} $\rightarrow$ \textit{Can't set min...}} [\inlineReviewLink{1149}]) and string value (\inlineReviewHref{MDI0OlB1bGxSZXF1ZXN0UmV2aWV3Q29tbWVudDUzMjQ1NjQzMA==}{\textit{aproveInboxItemById} $\rightarrow$ \textit{approveInboxItemById}} [\inlineReviewLink{1642}]). Additionally, developers changed \numberToBeChecked{five} direct string values to adhere to project standards 
[\inlineReviewLink{2318}] 
or fix values 
[\inlineReviewLink{39}] 
to alternatives that adhere to project standards.

\vspace{5pt}
\noindent\textbf{Inadequate logging and monitoring}. Developers have fixed concerns related to logging and exceptions in \numberToBeChecked{\waitingReview{15}} patches. In these patches, the developers added new logging instructions in \numberToBeChecked{\waitingReview{three}} instances to facilitate the understanding of the execution of a method or statement (e.g., \inlineReviewHref{MDI0OlB1bGxSZXF1ZXN0UmV2aWV3Q29tbWVudDQyMDc0NjczNw==}{adding logging after processing a list} [\inlineReviewLink{1135}]), and removed unnecessary logging in \numberToBeChecked{\waitingReview{six}} instances (e.g., \inlineReviewHref{MDI0OlB1bGxSZXF1ZXN0UmV2aWV3Q29tbWVudDQ0MTE0NDIyOA==}{removing log for local debugging with \texttt{System.out.println}} [\inlineReviewLink{1021}]). Moreover, developers changed logging levels (e.g., \inlineReviewHref{MDI0OlB1bGxSZXF1ZXN0UmV2aWV3Q29tbWVudDM3NDA1ODgzMg==}{from \textit{fine} to \textit{finest}} [\inlineReviewLink{1751}]) and simplified parameters of formatted text when logging (\inlineReviewHref{MDI0OlB1bGxSZXF1ZXN0UmV2aWV3Q29tbWVudDU5ODU0MTI5NQ==}{use direct value instead of variable} [\inlineReviewLink{1173}]). Additionally, in \inlineReviewHref{MDI0OlB1bGxSZXF1ZXN0UmV2aWV3Q29tbWVudDQ2MjMzMzMxNw==}{one case} [\inlineReviewLink{919}], a patch replaced 
\texttt{RuntimeException} by a more specific type, \texttt{EncodingException}.

\vspace{5pt}
\noindent\textbf{Missing constants}. 
In \numberToBeChecked{\waitingReview{six}} patches developers created new constants or used available ones to replace literals. They created new constants to replace literal arguments in \numberToBeChecked{\waitingReview{four}} instances (e.g., \inlineReviewHref{MDI0OlB1bGxSZXF1ZXN0UmV2aWV3Q29tbWVudDQxMzgxNTk3OA==}{\texttt{bodyParams.put("username", user)} $\rightarrow$ \texttt{bodyParams.put(PARAM\_USERNAME, user)}} [\inlineReviewLink{2163}]) and replaced literals by library constants in \numberToBeChecked{two} 
[\inlineReviewLink{590}]. 

\begin{summary}
\begin{small}
\textbf{Answer to RQ4.}
We found \numberToBeChecked{\waitingReview{323}} patches where developers added, removed, changed, replaced, and moved code elements to improve code understandability. The most common type of understandability improvement was modifying an identifier to express the meaning of a code element  (\numberToBeChecked{\waitingReview{50}}). This was followed by removing unused constants, imports, classes, methods, or variables (\numberToBeChecked{\waitingReview{19}}), changing text in existing Javadoc to improve grammar or fix a typo (\numberToBeChecked{\waitingReview{15}}), and removing duplicate code or processing (\numberToBeChecked{\waitingReview{13}}).

\vspace{5pt}
\noindent\textbf{Implications.} These results show that there is considerable variation in understandability improvements: some can be easily supported by linters (e.g., removing unused elements) whereas others require approaches that may be challenging even considering recent advances in machine learning (e.g., removing functionally equivalent code that is not textually similar). They emphasize how diverse tools can be employed to address the different understandability smells.
\end{small}
\end{summary}

\subsection{\rqfive (RQ5)}

We analyzed whether or not developers kept the \numberToBeChecked{\waitingReview{323}} patches analyzed in \autoref{results:rq4} throughout the history of the patched files. 
We observed that none of the patches implementing code understandability improvements were reverted in the pull requests where they were applied. However, we came across \numberToBeChecked{seven} instances where there are commits with reversions that were undone in subsequent commits of improvement patches in the same pull request. For example, in \inlineReviewHref{MDI0OlB1bGxSZXF1ZXN0UmV2aWV3Q29tbWVudDQ5OTg3OTMyOA==}{one case} [\inlineReviewLink{1471}] the improvement patch followed a suggestion to remove \texttt{@Deprecated}, but in the following commit this annotation was added again and, later, in subsequent commit, it was removed as suggested initially by reviewer. Also, we discover that \numberToBeChecked{\waitingReview{97.2\%}} of the improvement patches were successfully merged. The remaining patches (\numberToBeChecked{nine}) involved code snippets that were superseded by subsequent commits within the same pull request. For example,  \inlineReviewHref{MDI0OlB1bGxSZXF1ZXN0UmV2aWV3Q29tbWVudDQ4ODc0NzM2MQ==}{a patch added Javadoc to a method and later both the method and the associated Javadoc were removed} [\inlineReviewLink{1338}],  \inlineReviewHref{MDI0OlB1bGxSZXF1ZXN0UmV2aWV3Q29tbWVudDQ2NjU5OTAzOA==}{a method named \texttt{inAppDisplayhold} was renamed to \texttt{displayPaused} in a patch and later renamed to \texttt{autoDisplayPaused}} [\inlineReviewLink{1203}], \inlineReviewHref{MDI0OlB1bGxSZXF1ZXN0UmV2aWV3Q29tbWVudDQ1MTAwOTA3OA==}{or the patched file was removed entirely} [\inlineReviewLink{416}].

We also evaluate whether the (\numberToBeChecked{\waitingReview{314}}) patches implementing suggested understandability improvements and integrated into the codebase were reverted at some point in the history of the project after the original pull request and the associated code review were closed. Out of the total \numberToBeChecked{\waitingReview{314}} patches, \numberToBeChecked{\waitingReview{82.8\% (260/314)}} are present in the latest version of the codebase, although in \numberToBeChecked{20} cases, the developers renamed or moved the patched files. However, we cannot find the patch in the last version of the codebase in \numberToBeChecked{\waitingReview{54}} patches. Among these, the developers replaced the patch with an alternative solution in \numberToBeChecked{\waitingReview{ten}} observations. For instance, a subsequent patch changed a snippet \inlineReviewHref{MDI0OlB1bGxSZXF1ZXN0UmV2aWV3Q29tbWVudDUwMDU2MDg3Ng==}{with functional implementation (commit a0c1cbf9746ed9486f08ce4aa2376828e846c786) by an alternative imperative implementation} [\inlineReviewLink{1453}]. 
In \numberToBeChecked{\waitingReview{24}} patches, the developers removed the code snippet that contained the patch, e.g., \inlineReviewHref{MDI0OlB1bGxSZXF1ZXN0UmV2aWV3Q29tbWVudDQ0Njg1MjAwMA==}{the patch included the deprecated annotation and deprecated methods were deleted} [\inlineReviewLink{2263}]. Still, in \numberToBeChecked{\waitingReview{18}} patches, the developers deleted the entire patched file, e.g., \inlineReviewHref{MDI0OlB1bGxSZXF1ZXN0UmV2aWV3Q29tbWVudDQ2NjQzODUwNQ==}{AssemblyJsonServlet.java deleted} [\inlineReviewLink{1935}]. Finally, the patches were reverted and kept reverted until the latest version of the codebase in \numberToBeChecked{two} of these \numberToBeChecked{\waitingReview{54}} observations. One such case involved \inlineReviewHref{MDI0OlB1bGxSZXF1ZXN0UmV2aWV3Q29tbWVudDM5OTcxNTAzNg==}{the removal of an unnecessary class} [\inlineReviewLink{2073}], but the file was added back into the codebase and later renamed.

As a corollary result of this analysis, we observe that most of the analyzed patches implementing code understandability improvements \numberToBeChecked{\waitingReview{(97.2\%)}} were integrated into the codebase, whereas Kalliamvakou et al.~\cite{Kalliamvakou2014} reported only \numberToBeChecked{44\%} of pull requests in GitHub (in general) being merged. This stark difference may highlight the high acceptance rate of the reviewers' suggestions for code understandability improvements. It must be taken with a grain of salt, though, as the study of Kalliamvakou et al. was published in 2014. In addition, only two out of \numberToBeChecked{\waitingReview{323}} patches were reverted and maintained their reversion in the latest codebase version.

\begin{summary}
\begin{small}
\textbf{Answer to RQ5.}
After analyzing the history of \numberToBeChecked{\waitingReview{323}} patches, we discovered that \numberToBeChecked{\waitingReview{80.5\% (260 out 323)}} were present in the latest version of the codebase, while the remaining ones were either removed or replaced by alternative solutions. Additionally, we observed that \numberToBeChecked{none} of the patches were reverted during the pull request. \waitingReview{Finally, we found \numberToBeChecked{two} cases} that were removed later during the history of the codebase.

\vspace{5pt}
\noindent\textbf{Implications.} The results in this section suggest that understandability improvements tend to be stable, i.e., it is unlikely that an improvement will later disappear within a project. \waitingReview{Thus, training machine learning models to identify code understandability smells and improving their understandability, one of the goals of code reviews, may yield better results than automating code the whole spectrum of code review activities, including the identification of code with poor understandability and the refactoring of these instances into more readable code.~\cite{Tufano:2021:TAC,Lin:2023:TAC}}.
\end{small}
\end{summary}

\begin{table}[t]
	\caption{Out of the \numberToBeChecked{\waitingReview{323}} accepted patches addressing \numberToBeChecked{understandability smells}, the frequencies of patch merging, reversion within the pull request where it was accepted, permanence in the code based until the last version of the file, and reversion at any point in the history of the project.}\label{tab:rq5:result}.
	\centering
	\footnotesize
 \vspace{-5pt}
	\begin{tabular}{@{}p{0.6\columnwidth} r r@{}}
		\toprule
		Patch status (\# analyzed patches) & \multicolumn{1}{c}{Yes} & \multicolumn{1}{c}{No}\\
		\midrule
		Patch merged (323) & 314 (97.2\%) & 9 \hspace*{1.8pt} (2.8\%) \\ 
        Patch reverted in the pull request (9) & 0 \hspace*{7.8pt} (0\%) & 9 \hspace*{1.4pt} (100\%) \\
		Patch in the last version of the file (314) & 260 (82.8\%) & 54 (17.2\%) \\
		Patch reverted in the codebase history (54) & 2 \hspace*{2pt} (3.7\%) & 52 (96.3\%) \\
		\bottomrule
	\end{tabular}
 \vspace{-8pt}
\end{table}

\subsection{\rqsix (RQ6)}


We investigated the coverage of the \numberToBeChecked{code understandability smells} identified in our study by rules implemented in Spotbugs, PMD, SonarQube, and Checkstyle. To investigate if any of these linters' rules can flag the \numberToBeChecked{\waitingReview{323}} understandability smells presented in \autoref{results:rq4}, we manually checked each observation of an \numberToBeChecked{understandability smell} against each tool to determine if it could detect it. For example, for an observation of \textsc{Unnecessary code} related to the use of \texttt{import} clauses, we inspected the rules \textit{UnnecessaryImport}, \textit{Unnecessary imports should be removed}, and \textit{UnusedImports} in PMD, SonarQube, and Checkstyle respectively, which could detect the \textit{unused import} pointed out by the reviewer. There is no rule to detect this particular instance of \textsc{Unused code} in Spotbugs. We also considered occurrences where it may be necessary to configure a rule for the linter to detect it. For instance, the rule \textit{MissingJavadocMethod} of Checkstyle needs configuring the scope to warn about missing Javadoc in methods with ``package'' visibility (otherwise, it only reports issues with ``public'' ones). \autoref{tab:rq6:result} presents the number of occurrences of each understandability smell that each linter could detect. The rows represent each type of \numberToBeChecked{understandability smell}, while the columns correspond to each linter. The column \textit{Coverage} presents the number of instances of each smell that can be identified by at least one linter.

\begin{table*}[t]
        \vspace*{-10pt}
	\caption{The frequency of understandability smells that are covered by linters.}\label{tab:rq6:result}
	\centering
	\footnotesize
 \vspace*{-8pt}
	\begin{tabular}{@{}lrrrrr}
		\toprule
		Understandability smells (\# occurrences) & 
            Spotbugs &
            PMD &
            SonarQube &
            Checkstyle &
            Coverage \\
		\midrule
	    \textsc{\scriptsize{Incomplete or inadequate code documentation (71)}} &
                0 \hspace*{1.8pt} (\tablecellcolorRQ6{0.0}\%) & 10 (\tablecellcolorRQ6{14.1}\%) & 7 \hspace*{1.8pt} (\tablecellcolorRQ6{9.9}\%) & 14 (\tablecellcolorRQ6{19.7}\%) & 14 (\tablecellcolorRQ6{19.7}\%)  \\
            \textsc{\scriptsize{Bad identifier (71)}} &
                1 \hspace*{1.8pt} (\tablecellcolorRQ6{1.4}\%) & 6 \hspace*{1.8pt} (\tablecellcolorRQ6{8.5}\%) & 3 \hspace*{1.8pt} (\tablecellcolorRQ6{4.2}\%) & 5 \hspace*{1.8pt} (\tablecellcolorRQ6{7.0}\%) & 7 \hspace*{1.8pt} (\tablecellcolorRQ6{9.9}\%)  \\
            \textsc{\scriptsize{Complex, long or inadequate logic (54)}} &
                0 \hspace*{1.8pt} (\tablecellcolorRQ6{0.0}\%) & 8 (\tablecellcolorRQ6{14.8}\%) & 11 (\tablecellcolorRQ6{20.4}\%) & 7 (\tablecellcolorRQ6{13.0}\%) & 16 (\tablecellcolorRQ6{29.6}\%)  \\
	    \textsc{\scriptsize{Unnecessary Code (43)}} &
                6 (\tablecellcolorRQ6{14.0}\%) & 11 (\tablecellcolorRQ6{25.6}\%) & 18 (\tablecellcolorRQ6{41.9}\%) & 7 (\tablecellcolorRQ6{16.3}\%) & 22 (\tablecellcolorRQ6{51.2}\%)  \\
            \textsc{\scriptsize{Inconsistent or disrupted formatting (37)}} &
                0 \hspace*{1.8pt} (\tablecellcolorRQ6{0.0}\%) & 3 \hspace*{1.8pt} (\tablecellcolorRQ6{8.1}\%) & 14 (\tablecellcolorRQ6{37.8}\%) & 25 (\tablecellcolorRQ6{67.6}\%) & 25 (\tablecellcolorRQ6{67.6}\%)  \\
            \textsc{\scriptsize{Wrong, missing, or inadequate string expression or literal (26)}} &
                0 \hspace*{1.8pt} (\tablecellcolorRQ6{0.0}\%) & 0 \hspace*{1.8pt} (\tablecellcolorRQ6{0.0}\%) & 0 \hspace*{1.8pt} (\tablecellcolorRQ6{0.0}\%) & 0 \hspace*{1.8pt} (\tablecellcolorRQ6{0.0}\%) & 0 \hspace*{1.8pt} (\tablecellcolorRQ6{0.0}\%)  \\
            \textsc{\scriptsize{Inadequate logging and monitoring (15)}} &
                0 \hspace*{1.8pt} (\tablecellcolorRQ6{0.0}\%) & 2 (\tablecellcolorRQ6{13.3}\%) & 2 (\tablecellcolorRQ6{13.3}\%) & 0 \hspace*{1.8pt} (\tablecellcolorRQ6{0.0}\%) & 2 (\tablecellcolorRQ6{13.3}\%)  \\
            \textsc{\scriptsize{Missing constant usage (6)}} &
                0 \hspace*{1.8pt} (\tablecellcolorRQ6{0.0}\%) & 0 \hspace*{1.8pt} (\tablecellcolorRQ6{0.0}\%) & 3 (\tablecellcolorRQ6{50.0}\%) & 3 (\tablecellcolorRQ6{50.0}\%) & 3 (\tablecellcolorRQ6{50.0}\%)  \\
        \midrule
            Total (323) &
                7 \hspace*{1.8pt} (\tablecellcolorRQ6{2.2}\%) & 40 (\tablecellcolorRQ6{12.4}\%) & 58 (\tablecellcolorRQ6{18.0}\%) & 61 (\tablecellcolorRQ6{18.9}\%) & 89 (\tablecellcolorRQ6{27.6}\%)  \\
        \bottomrule
	\end{tabular}
        \vspace*{-13pt}
\end{table*}

Our analysis indicates that \numberToBeChecked{\waitingReview{89 out of 323}} occurrences of \numberToBeChecked{understandability smells} can be detected by any of the linters. This highlights wasted human effort, as automated tools could be pointing these issues out. 
On the one hand, Checkstyle and Sonarqube are responsible for detecting \numberToBeChecked{more than half} of detectable occurrences. On the other hand, Spotbugs detects only  \numberToBeChecked{seven} occurrences overall, where \numberToBeChecked{six} are related to \textsc{unnecessary code}. Furthermore, we did not find rules that could detect the occurrences related to \textsc{Wrong, missing, or inadequate string expression or literal}, either because the tools do not inspect string literals or because they are natural language text.

The \numberToBeChecked{understandability smell} that can be more often detected in our dataset by the linters is \textsc{Inconsistent or disrupted formatting}. Rules such as \textit{NeedBraces} (Checkstyle), \textit{Lines should not be too long} (SonarQube), and \textit{UselessParentheses} (PMD) can detect the majority (\numberToBeChecked{\waitingReview{67.6\%}}) of the occurrences of this smell. Additionally, these tools acknowledge that different projects have different styles. For example, in a code review comment, the reviewer suggested that \inlineReviewHref{MDI0OlB1bGxSZXF1ZXN0UmV2aWV3Q29tbWVudDQ2MTc1OTE2Mw==}{\textit{``bracket [left curly brace] should be on the next line''}} [\inlineReviewLink{1487}], i.e., the left curly brace should be in beginning a new line instead of the end of current statement line. Accounting for the possibility of some projects using left curly braces at the end of the current line, SonarQube contains rules for both cases, \textit{An open curly brace should be located at the beginning of a line} and \textit{An open curly brace should be located at the end of a line}.

The two most popular \numberToBeChecked{understandability smells}, \textsc{Bad identifier} and \textsc{Incomplete or inadequate code documentation}, are the second and fourth less detectable by linters, respectively. The majority of the occurrences of these \numberToBeChecked{understandability smells} are associated with natural language text (e.g., typo in identifier and issue in content of Javadoc). The cases of \textsc{Bad identifier} detected by the linters are those related to style (e.g., \inlineReviewHref{MDI0OlB1bGxSZXF1ZXN0UmV2aWV3Q29tbWVudDQ0NTgyNzc0OQ==}{parameters starting with capital letter} [\inlineReviewLink{579}]). In these cases, rules such as \textit{FormalParameterNamingConventions} (PMD), Nm: \textit{Field names should start with a lower case letter (NM\_FIELD\_NAMING\_CONVENTION)} (Spotbugs), and \textit{LocalVariableName} (Checkstyle) can be used to detect them. In the case of \textsc{Incomplete or inadequate code documentation}, there are only rules for missing documentation, e.g., \textit{CommentRequired} (PMD); missing or incorrect order of tags in Javadoc, e.g., \textit{WriteTag} (Checkstyle); and TODO code comments, e.g., \textit{Track uses of ``TODO'' tags} (SonarQube). Many instances of these smells require an in-depth understanding of the associated elements, e.g., to produce an identifier that better represents the purpose of a method or to improve an explanation about an element. At the same time, the most common change applied to Javadoc documentation in our study is fixing typos (\autoref{tab:rq4:result}), which could be automated.  

Among the remaining \numberToBeChecked{\waitingReview{234 out of 323}} instances of \numberToBeChecked{understandability smells}, we identify occurrences that could be covered by the investigated linters by performing small adaptations to existing rules. For example, the linters could cover the \textit{missing parentheses} (e.g., \inlineReviewLink{1453}) by creating an inverse rule to the \textit{UselessParentheses} rule of PMD or ``\textit{UnnecessaryParentheses}'' rule of Checkstyle. This new rule could suggest the inclusion of additional parentheses in boolean expressions to highlight evaluation order. Additionally, the linters could identify opportunities to use primitive types instead of the corresponding boxed types (e.g., \inlineReviewLink{1201}). As another example, the linters could detect when an exception is created without a message (e.g., \inlineReviewLink{1345}). However, other instances of \numberToBeChecked{understandability smells} would require more sophisticated analyses. For example, to detect the occurrences of incomplete documentation, unnecessary logging messages, confusing string messages, inadequate use of method call chains, and missing logging, it would be necessary to understand the context of the source code, the culture of the project, the intention of the developers, predict the system behavior at run time, or to work with both Java and natural languages.  

\begin{summary}
\begin{small}
\textbf{Answer to RQ6.}
We found \numberToBeChecked{89 out of 323} occurrences of \numberToBeChecked{understandability smells} that could be detected by the linters. Some of the remaining \numberToBeChecked{234} occurrences of \numberToBeChecked{understandability smells} could be detected by creating new rules in these linters, while other occurrences require a deeper and context-dependent analysis that goes beyond what linters typically do.

\vspace{5pt}
\noindent\textbf{Implications.} The four linters combined have rules to cover less than \numberToBeChecked{30\%} of the instances of understandability smells from our dataset. This suggests that developers waste effort during code reviews by alluding to issues that existing automated tools could detect. Among the cases that the linters cannot detect, many of them require an in-depth ``understanding'' of the program under analysis or the context in which it is inserted. In particular, current linters are unable to cope with code understandability issues related to natural language. Some of these cases present opportunities to devise new tools, e.g., to better summarize the intent of a code element or to give it a more illustrative name. Others could be addressed by simple improvements to current linters, e.g., pointing out typos in documentation, identifiers, and string literals. This also applies to rules related to other aspects, such as messages in exceptions and the use (or not) of boxed types. 
\end{small}
\end{summary}

\section{Threats to validity}

In this section, we discuss the threats to the validity of our study.

\vspace{5pt}
\noindent\textbf{Selection of projects.}
We constructed our dataset by mining software repositories from GitHub. 
However, researchers must be cautious about potential pitfalls~\cite{Kalliamvakou2014}. To ensure data quality, we applied filters to exclude personal projects, repositories with small contributor groups, and forked repositories. Moreover, we utilized a heuristic to select projects with an active code review process during a specific period. Also, we manually analyzed each remaining repository to guarantee the relevance of the dataset, excluding non-software repositories. Additionally, some projects may have configured an ASAT (Automated Static Analysis Tool) to prevent certain issues related to code understandability from being introduced into the codebase. However, since we did not assess whether any project adopted this approach, it is possible that some understandability smells may not have been identified by our study.

\vspace{5pt}
\noindent\textbf{Size of the sample of code review comments.}
Before analyzing our sample of \numberToBeChecked{2,401} code review comments to determine whether they are about code understandability improvements, we 
explored various natural language processing techniques, such as LDA, GSDMM, LSA, and LDA Guided, to extract topics and create clusters for both understandability and non-understandability improvements. Unfortunately, we encountered challenges in obtaining satisfactory results due to two main reasons: (i) our training sample was relatively small, both in terms of the number of comment instances and the length of their text, and (ii) there were significant similarities between comments related and unrelated to code understandability. 
This problem has been reported by previous work as well~\cite{Ebert:2017:CDC, tufan2021towards}.

\vspace{5pt}
\noindent\textbf{Manual analysis and subjectivity.}
We conducted a study where we manually evaluated \numberToBeChecked{2,401} code review comments. 
Manual categorization may result in bias due to different interpretations by human coders. 
Some comments that were analyzed lack precision and clarity. 
This could affect the validity of any analysis or decision-making based on the text. 
Additionally, it is possible that there is overlap between different smells. In these cases, we analyze the reviewers' comments to understand the intent of the proposed change. 
Each comment was independently classified by at least two authors, and whenever a disagreement was encountered, up to five discussed the classification. We also held sessions to discuss dubious comments and to align the classification process.
Furthermore, we analyzed the source code to which each comment refers and subsequent changes to it that may have been triggered by the comment.

We also conducted a manual analysis of rules in four linters that could potentially detect understandability smells highlighted by reviewers. For each instance of an understandability smell that we analyzed, we searched through a set of 1,315 rules to find the ones that could identify the smell. Some of these rules can be configured to detect a smell, and we considered all possible configurations. However, we decided not to include rules that relied on regex patterns due to their specificity. It is important to note that there may be some rules that we missed during this process, which could result in some issues being classified as not covered by linters.


\vspace{5pt}
\noindent\textbf{Code review bots.} During the code review process for pull requests, we came across bots that inspect the source code, i.e., bots that work as linters. These bots, like \texttt{vaadin-bot} and \texttt{codeclimate}, identify potential issues based on pre-configured rules. For instance, \texttt{vaadin-bot}\footnote{\url{https://github.com/vaadin/flow/pull/8409\#discussion_r429865662}} found a deprecated method annotation and suggested removing it in the future, while \texttt{codeclimate}\footnote{\url{https://github.com/triplea-game/triplea/pull/7241\#discussion_r460364607}} commented on a method that exceeded the allowed number of lines and suggested refactoring. \texttt{Codacy}, \texttt{Lint-staged}, and \texttt{Prettier} are examples of popular bots that review the source code searching for violations of their pre-configured rules. We decided not to investigate these bots specifically, considering that their rules are similar to the linters studied previously. For example, codeclimate leverages SonarJava\footnote{\url{https://docs.codeclimate.com/docs/sonar-java}}, also used by SonarQube, to verify Java programs.

\vspace{5pt}
\noindent\textbf{External Validity.} This study focused exclusively on open-source projects, and as such, the findings may not be fully representative of all software development contexts. To address this concern, we made efforts to include a diverse range of software projects in our analysis. Nonetheless, it is essential to acknowledge that the identified code understandability smells and corresponding understandability improvements might not be universally applicable across all programming languages, frameworks, development environments, or development teams. Furthermore, a limitation inherent to using GitHub is that many engineered projects in the platform are mostly developer-centric, e.g., libraries, frameworks, etc. This means that many sub-domains, such as games and mobile apps, are under-represented.
Additionally, our study focused on code reviews conducted in 2020, which may have been impacted by COVID-related issues. Therefore, it is unclear how much we can generalize the findings.
In our analysis, we did not take into account the experience level of the developers and reviewers. This means that we may have factored in the opinions of inexperienced individuals. However, we carefully selected a diverse sample group that included many developers and reviewers. This helped to minimize the impact of this issue.

\section{Related Work}
\label{sec:related_works}


In this section, we examine related work. We start by contrasting the understandability smells and improvements we have identified in practice with recommendations and experimental results reported in the literature (\autoref{sec:lit_prac}). This helps us identify mismatches and accordance between research and practice, as well as research gaps. We then review previous work discussing models to assess code understandability (\autoref{sec:understand_studies}), following up with papers that have investigated manifestations of confusion and reactions to lack of necessary information during code reviews (\autoref{sec:reviews}). 

\subsection{Coding alternatives for understandability: literature vs. practice}\label{sec:lit_prac}

We found in the literature many studies about code understandability. Among these studies, the researchers investigated the best code alternatives for code understandability. Oliveira et al.~\cite{Oliveira2023} conducted a systematic literature review that categorized and organized studies that used empirical evaluation to compare code alternatives for formatting elements. For instance, Gopstein et al.~\cite{Gopstein2017}, Medeiros et al.~\cite{Medeiros2019}, and Langhout and Aniche~\cite{langhout2021} examined the impact of explicit block delimiters (i.e., curly braces in Java) compared to the omission of block delimiters when possible. Furthermore, we found studies that compared other code elements, such as monolithic expressions compared to multiple shorter expressions~\cite{dosSantos2018}. We aim to compare our findings from practical experience with what we gathered from these studies. 

Our focus in this paper is on understandability improvements, i.e., situations where existing code exhibits an understandability smell and is replaced by an alternative solution that is more understandable. Thus, in this section, we focus on the five \numberToBeChecked{understandability 
smells} where, to the best of our knowledge,  alternative solutions were compared in existing studies from a human-centric perspective, e.g., based on the preference or performance of study participants: \textsc{Bad identifier}, \textsc{Unnecessary code}, \textsc{Inconsistent or disrupted formatting}, \textsc{Complex, long, or inadequate logic}, and \textsc{Missing constant usage}. 

\vspace{5pt}
\noindent\textbf{\textsc{Bad identifier}}. Chaudhary and Sahasrabuddhe \cite{Chaudhary1980} evaluated the impact on code understandability of using identifier names unrelated to the program domain. This study concluded that the identifier names with meaning related to the problem domain facilitate understanding. In this direction, the practitioners improve the identifier names to better match their \inlineReviewHref{MDI0OlB1bGxSZXF1ZXN0UmV2aWV3Q29tbWVudDUwODIxMzQ4Mw==}{intent} and \inlineReviewHref{MDI0OlB1bGxSZXF1ZXN0UmV2aWV3Q29tbWVudDUxNzMwNTQxNw==}{type}. Santos and Gerosa~\cite{dosSantos2018} compared using fully-qualified names and \texttt{import} clauses with shorter type names. They found out that participants preferred to avoid fully-qualified names and add \texttt{import} clauses, which is what the \inlineReviewHref{MDI0OlB1bGxSZXF1ZXN0UmV2aWV3Q29tbWVudDQ5OTIyNzA3OQ==}{practitioners in our study prefer}. Schankin et al.~\cite{Schankin2018} investigated whether a more descriptive name (e.g., \textit{singleWordParts}) is easier to read than a single-word one (e.g., \textit{parts}). They found that a descriptive name was the best in one scenario, and there was no difference in the other. The practitioners also considered improving understandability with \inlineReviewHref{MDI0OlB1bGxSZXF1ZXN0UmV2aWV3Q29tbWVudDQ1MzEyOTQ2OA==}{descriptive names}. Sharif and Maletic~\cite{Sharif2010} and Binkley et al.~\cite{Binkley2013} investigated the effect of the style of the identifier (underscore and camel case) on code understandability. Sharif and Maletic found that the use of an underscore to separate parts of an identifier improves understandability, whereas Binkley et al. found out that camel case works better. For practitioners in Java, using the \inlineReviewHref{MDI0OlB1bGxSZXF1ZXN0UmV2aWV3Q29tbWVudDYyNzA3OTUwNg==}{camel case style} for non-constant identifiers improves code understandability.

\vspace{5pt}
\noindent\textbf{\textsc{Unnecessary code}}. Gopstein et al.~\cite{Gopstein2017} and Torres et al.~\cite{Torres:2023:ICC} investigated whether the presence of dead, unreachable, or repeated code (e.g., \texttt{v1 = 1; v1 = 2}), leads to misunderstanding in C and JavaScript programs, respectively. Both studies found that the presence or absence of this pattern did not significantly impact code understandability. In contrast, practitioners seem to believe this is an important issue in practice: \numberToBeChecked{13.3\%} of all the \numberToBeChecked{323} analyzed code review comments pertain to \textsc{unnecessary code}. The practitioners removed the \inlineReviewHref{MDI0OlB1bGxSZXF1ZXN0UmV2aWV3Q29tbWVudDQ1OTk4NTYwMw==}{unnecessary code}, \inlineReviewHref{MDI0OlB1bGxSZXF1ZXN0UmV2aWV3Q29tbWVudDU0Njc4NDIyOQ==}{commented out source code}, and \inlineReviewHref{MDI0OlB1bGxSZXF1ZXN0UmV2aWV3Q29tbWVudDQ2ODA4ODUyMQ==}{duplicated code} to improve code understandability.

\vspace{5pt}
\noindent\textbf{\textsc{Inconsistent or disrupted formatting}}. Santos and Gerosa~\cite{dosSantos2018}, and Sampaio and Barbosa~\cite{sampaio2016} evaluated whether different conventions in terms of vertical and horizontal space between the code elements impact code understandability and did not find significant differences. Conversely, practitioners consider that spacing is an issue worth raising during code review. Out of the \numberToBeChecked{\waitingReview{323}} analyzed comments, \numberToBeChecked{\waitingReview{26}} (\numberToBeChecked{\waitingReview{8\%}}) pertain to spacing. Sometimes, they \inlineReviewHref{MDI0OlB1bGxSZXF1ZXN0UmV2aWV3Q29tbWVudDU0MzQ5ODc3Mg==}{remove space}, and others \inlineReviewHref{MDI0OlB1bGxSZXF1ZXN0UmV2aWV3Q29tbWVudDQyOTQ3OTQwNA==}{add space}. Sykes et al.~\cite{Sykes1983}, Sampaio and Barbosa~\cite{sampaio2016}, Gopstein et al.~\cite{Gopstein2017}, Langhout and Aniche~\cite{langhout2021}, Torres et al.~\cite{Torres:2023:ICC}, and Medeiros et al.~\cite{Medeiros2019} investigated the presence or omission of the block delimiter (e.g., curly braces in Java) in blocks with one statement. Gopstein et al., Torres et al., Langhout and Aniche, and Medeiros et al. found that the presence of the block delimiter improves code understandability (in C, JavaScript, and Java). The other researchers did not find any difference. 
The practitioners preferred \inlineReviewHref{MDI0OlB1bGxSZXF1ZXN0UmV2aWV3Q29tbWVudDM5NTgwMjc1OQ==}{to use block delimiters even in a one-statement block}. Geffen and Maoz~\cite{Geffen2016} investigated the effect of ordering methods in understanding the code. For some cases ordering considering the order of calling and the connectivity between the methods is the best for code understandability. The practitioners consider that \inlineReviewHref{MDI0OlB1bGxSZXF1ZXN0UmV2aWV3Q29tbWVudDUzNDE0NDYzMg==}{the calling order} improves code understandability.

\vspace{5pt}
\noindent\textbf{\textsc{Complex, long, or inadequate logic}}. 
Gopstein et al.~\cite{Gopstein2017}, Medeiros et al.~\cite{Medeiros2019}, Torres et al.~\cite{Torres:2023:ICC}, Costa et al.~\cite{Costa:2023:SCT}, and Langhout and Aniche~\cite{langhout2021} investigated which one promotes better understandability: the conditional operator or \texttt{if} statements, in C, JavaScript, Python, and Java small programs. Gopstein et al. and Medeiros et al. found that \texttt{if} statements improve code understandability (in C programs) whereas Langhout and Aniche, Costa et al., and Torres et al. did not find a statistically significant difference. The practitioners considered both \inlineReviewHref{MDI0OlB1bGxSZXF1ZXN0UmV2aWV3Q29tbWVudDQwOTAwMjcwOQ==}{\texttt{if} statement} and \inlineReviewHref{MDI0OlB1bGxSZXF1ZXN0UmV2aWV3Q29tbWVudDQ4OTA4Mzk1Nw==}{ternary operator} to improve code understandability in distinct scenarios. 
Lucas et al.~\cite{Lucas:2019:DIL} compared the use of lambdas and anonymous inner classes based on the opinions of 28 experienced Java developers. Their results are aligned with our findings when examining code review comments: practitioners prefer lambda expressions.  
Most of the instances of the \textsc{Complex, long, or inadequate} understandability smell are either too specific to be studied in a more general context, e.g., object creation that can be avoided by using an existing attribute, or, to the best of our knowledge, have not been studied, e.g., using \texttt{try} statements that manage resources, instead of \texttt{try} with resources.

\vspace{5pt}
\noindent\textbf{\textsc{Missing constant usage}}. Gopstein et al.~\cite{Gopstein2017}, Langhout and Aniche~\cite{langhout2021}, and Torres et al.~\cite{Torres:2023:ICC} evaluated the effect of using constant variables instead of direct values on code understandability. They did not find a statistically significant difference. The practitioners \inlineReviewHref{MDI0OlB1bGxSZXF1ZXN0UmV2aWV3Q29tbWVudDQxMzgxNTk3OA==}{created and used constants instead of direct values} to improve code understandability.

\subsection{Estimating and classifying code understandability
}\label{sec:understand_studies}



Piantadosi et al.~\cite{piantadosi2020} investigated the frequency, methods, and reasons for changes in code understandability \waitingReview{(\textit{code readability} in their paper)} during the evolution of open-source software. 
They built a model to define states (non-existing, other-name, readable, and unreadable) in which a file can be in at a given time. Using this model, they analyzed the transitions related to code understandability in the commit history of 25 open-source software projects. By analyzing more than 340,000 understandability transitions, they discovered that code understandability rarely changes during the evolution of the project. In a manual analysis of 57  understandability transitions, they found that improving or decreasing code understandability is mostly unintentional. 

Fakhoury et al.~\cite{Fakhoury2019} investigated whether state-of-the-art understandability models are able to capture \waitingReview{understandability} \waitingReview{(\textit{readability} in their paper)} improvements as explicitly tagged by open source developers in commit messages. The authors analyzed 548 commits from 63 engineered Java projects that explicitly mention understandability improvements in the commit message. They then used three state-of-the-art \waitingReview{understandability} models to measure how much these commits affected code \waitingReview{understandability}. They found that the models did not capture \waitingReview{understandability} improvements in most cases. Additionally, they used static analysis tools to investigate what type of understandability changes were applied in the commits related and unrelated to code understandability. They discovered that the former frequently fix concerns related to imports, white spaces, and braces, whereas the latter tends to introduce them.

Roy et al.~\cite{roy2020} proposed a model to detect code understandability \waitingReview{(\textit{readability} in their paper)} improvements made by developers in real-world scenarios using commit histories from open-source software projects by extending the dataset of 
Fakhoury et al.~\cite{Fakhoury2019}. They also included commits unrelated to code understandability improvements to create an oracle. The authors employed seven source code analysis tools to collect metrics before and after each commit at the file level. By comparing these metrics, they identified key differences that served as features for their model. The model achieved a precision of 79.2\% and a recall of 67\% on the test set.

Buse and Weimer~\cite{buse2010} proposed a method for measuring code understandability \waitingReview{(\textit{code readability} in their paper)} based on human notions of understandability. They collected feedback from 120 human annotators on their understanding of 100 code snippets, correlating this feedback with source code features. This data was then used to train a machine learning algorithm to classify code understandability. The resulting model predicted understandability judgments with an 80\% success rate. Posnett et al.~\cite{posnett2011} proposed a simplification of Buse and Weimer's~\cite{buse2010} model. They leverage source code size metrics and Halstead metrics~\cite{Halstead1977}. Using the dataset of Buse and Weimer~\cite{buse2010} to train and evaluate their model, they obtained better results using just three features in small code snippets: volume, entropy, and number of lines.

Scalabrino et al.~\cite{scalabrino2019} analyzed the correlation between 121 metrics related to code, documentation, and developers and six different proxies of code understandability, based on 444 human evaluations. The study revealed that none of the individual metrics strongly correlated with the proxies for code understandability. However, combining multiple state-of-the-art metrics resulted in models that showed a slight improvement in the regression performance of two proxies, Timed Actual Understandability (Correlation: +0.07; Mean absolute error: -0.02) and Actual Understandability (Correlation: +0.02; Mean absolute error: +0.00).

Lavazza et al.~\cite{lavazza2023} evaluated whether code measures are correlated with code understandability. They conducted an empirical study with students who performed maintenance tasks. The results showed that a code comprehensibility model cannot rely solely on measures of code structure. The obtained models were not very accurate, with the average prediction error being around 30\%.

These studies~\cite{buse2010,posnett2011,scalabrino2019, lavazza2023} have introduced models to quantitatively estimate the degree of code understandability or to detect code understandability improvements~\cite{piantadosi2020,roy2020}. In contrast, our work aims to delve into the code constructs and practices that enhance code understandability, specifically exploring code alternatives suggested by code quality specialists during software development. Additionally, we manually identified code understandability changes using code review comments rather than relying on an automatic approach based on metrics or keywords~\cite{piantadosi2020,Fakhoury2019,roy2020}. Furthermore, we categorized code understandability smells based on the associated issues and solutions instead of categorizing them in terms of software maintenance types~\cite{piantadosi2020} or rules of static analysis tools~\cite{Fakhoury2019}. Finally, our work acknowledges that different projects have different rules, priorities, and development cultures and these differences have an impact on what constitutes readable code. 

Some other studies \cite{MI201834, MI201860, MI2023} have proposed techniques for automating assessment through advanced methods, including convolutional neural networks. Mi et al. \cite{MI201834} introduced IncepCRM, a model that autonomously extracts multi-scale features from code, aided by human annotations to enhance accuracy. In a distinct approach \cite{MI201860}, an 83.8\% accuracy is achieved, surpassing prior machine-learning models. This is realized by transforming source code into matrices for convolutional neural networks, culminating in DeepCRM, an architecture of three networks trained on diverse preprocessed data. Similarly, in another study \cite{MI2023}, a graph-based method is adopted, parsing source code into a graph containing abstract syntax tree (AST), along with control and data flow edges, preserving semantic structural insights. This approach attains 72.5\% and 88\% accuracy in three-class and two-class classifications, respectively.
Our analysis of inline code reviews, focusing on developers' perspectives regarding code understandability, provides valuable insights for the referenced works. The research conducted by Mi et al. \cite{MI201834} could further benefit by incorporating insights from our study to refine their IncepCRM model. Our findings might aid in identifying specific aspects of code \waitingReview{understandability} highlighted during reviews, enabling more effective feature extraction for multi-scale representation. Similarly, the strategy proposed by the author in \cite{MI201860} could gain from our insights by considering the nuances of developer concerns about code understandability in the design of novel representation strategies. 
Integrating our unique dataset and analytical framework could enhance these approaches, contributing to advancements in code \waitingReview{understandability} evaluation.

\waitingReview{Vitale et al. \cite{vitale2023using} introduced an approach to improve code understandability (\textit{code readability} in their paper). They analyzed the revision history to identify commits aimed at improving understandability and collected a dataset of 122k such commits. They used the T5 model to emulate developers' actions in enhancing code understandability. Their findings showed that their approach accurately identified understandability-improving commits around 86\% of the time. However, the model's precision in suggesting actions to improve code understandability ranged between 21\% and 28\%.
Our findings lay the groundwork for the development of such predictive models. For example, we identified understandability smells that are currently not tackled by existing linters and therefore require more powerful approaches such as the use of LLMs. Furthermore, the dataset of understandability smells and improvements we have produced can be leveraged to fine-tune future language models.}

More generic LLM models could leverage our insights to enhance their predictive capabilities. These models could refine their feature extraction processes by considering the specific factors that developers prioritize for code understandability. Incorporating our findings into their frameworks could result in more accurate assessments of code \waitingReview{understandability}, aligning more closely with real-world developer concerns. Thus, our work offers a valuable resource for improving the effectiveness of broader LLM-based approaches to code evaluation. CodeBERT, introduced in \cite{feng-etal-2020-codebert} by Zhangyin Feng et al., is designed for tasks like code summarization and code-to-text generation. Analyzing inline reviews for \waitingReview{understandability} can offer valuable insights into the patterns and priorities developers have regarding code \waitingReview{understandability}. By integrating these insights, CodeBERT's performance can be enhanced, especially in generating human-readable code summaries and comments. Understanding the \waitingReview{understandability} aspects developers value most can enable models like CodeBERT to align more closely with human preferences in code generation and summarization.

\subsection{Understandability and Code Reviews}\label{sec:reviews}

In this section, we present studies that examine code review comments and how developers express confusion in these comments while reading code. 
The work by Dantas et al.~\cite{Dantas2023} is arguably the one that is the most related to our study. They evaluated code \waitingReview{understandability (\textit{code readability} in their paper)} improvements in the context of pull requests (PRs) and compared them with the rules of SonarQube. Dantas and colleagues utilized keywords such as ``readability'' and ``understandability'' and analyzed the context of proposed code changes to identify PRs that enhance code \waitingReview{understandability}. The researchers identified and categorized 284 PRs and found 26 different types of code \waitingReview{understandability} improvements. Among the 284 code \waitingReview{understandability} improvements classified, they found that 26 of them are detected by SonarQube. Differently, our study examined \numberToBeChecked{2,401} code review comments that suggested fine-grained source code improvements. We manually identified \numberToBeChecked{1,012} code review comments that suggested code understandability improvements based on the reviewers' intent. We classified the understandability smells and their solutions, as well as investigated the prevalence of the patches applied to improve code understandability. We have also analyzed whether understandability improvements are accepted or not and whether they are reversed. Finally, we analyzed the potential of four different linters to detect the understandability smells. 

The work of Pascarella et al.~\cite{pascarella2018} aimed to understand the information necessary for reviewers to conduct a proper code review. It suggests ways in which research and tool support can enhance the effectiveness and efficiency of developers in their role as reviewers. The study analyzed 900 code review comments from three large open-source projects and identified seven primary information needs of reviewers, such as understanding the uses of methods and variables declared or modified within the code. 
In a similar vein, Ebert et al.~\cite{Ebert:EMSE:2021} investigated the causes and consequences of confusion in code reviews and how developers typically handle such issues. The study involved a survey of 54 valid responses and a manual evaluation of 307 code review comments (both inline and general). The authors identified 30 sources of confusion, 14 ways that confusion impacts code review, and 13 coping strategies to deal with confusion in code reviews. The research revealed that developers often experience confusion during code reviews when dealing with long or complex code changes and changes that address multiple issues.

Our work differs from previous studies because we aim to evaluate how developers improve understandability through code review comments. Furthermore, the findings of our study can be used as recommendations to developers to improve their source code. This can lead to code that causes less confusion when being reviewed.

Other studies have explored approaches to automate code review, in the generation of patches~\cite{tufano2021,tufano2022} and of code review comments~\cite{Hong2022}. The former approach leverages LLMs and the latter employs information retrieval. Neither of them attempts to distinguish the purposes of the code review comments. Likewise, the study by Fregnan et al.~\cite{fregnan2022} evaluated what extent a machine learning-based technique can automatically classify review changes. The researchers classified 1,504 review changes manually and then evaluated three different machine learning algorithms to see if they could classify review changes automatically at two different levels of detail. Their findings indicate that machine learning can be effectively used to classify review-induced changes. This paper is complementary to these approaches in that it narrows down the scope of investigation to code understandability improvements.
It shows the relevance of code review comments aiming to improve understandability, highlights the high-level problems that these comments address, shows how many of the corresponding improvements materialize, and indicates that they are often accepted and rarely reversed. It also provides evidence that automating the application of improvements involves a variety of issues, targets different languages, natural and artificial, and in some cases can be solved easily whereas in others requires a deeper understanding.

\section{Conclusion}

Code understandability is crucial for efficient software development and maintenance, as it reduces the time and cost required for these processes. Despite its importance, improving code understandability is not straightforward, as the factors influencing it are not fully understood. Previous studies comparing different coding approaches have yielded inconclusive or controversial results. Additionally, style guides often do not align with experimental findings, leading to differing perceptions among developers.

To address this research gap, this study focused on code review comments in open-source projects, where developers actively strive to improve code understandability. The analysis of \numberToBeChecked{2,401} code review comments identified \numberToBeChecked{1,012} comments related to code understandability improvements. This highlights the significance of code review in enhancing code understandability. A deeper analysis of \numberToBeChecked{\waitingReview{385}} code review comments identified \numberToBeChecked{eight} categories of code understandability smells, such as \textsc{incomplete or inadequate code documentation} and \textsc{bad identifiers}. Developers often accepted the suggested improvements and rarely reverted them. The study also explored the types of patches applied to address code understandability smells, where \numberToBeChecked{\waitingReview{97.2\%}} of applied patches were integrated into the project's codebase. Additionally, while some of these issues can be readily detected and fixed by linters, others demand more intricate analysis, considering natural language components and project context.

This study contributes to the understanding of developers' concerns and actions regarding code understandability during code reviews. The dataset of code review comments serves as a valuable resource for researchers and developers, enabling the creation of automated tools to detect and repair code understandability smells. Such tools would allow developers to focus on other critical aspects of code review, promoting correctness and identifying security vulnerabilities.


\section*{Acknowledgements}
This work was partially supported by INES (www.ines.org.br), CNPq grant 465614/2014-0, FACEPE grants APQ-0399-1.03/17 and APQ/0388-1.03/14, CAPES grant 88887.136410/2017-00, the Swedish Foundation for Strategic Research (SSF), and the Wallenberg Artificial Intelligence, Autonomous Systems and Software Program (WASP) funded by Knut and Alice Wallenberg Foundation.

\bibliographystyle{IEEEtran}
\bibliography{references}

\end{document}